\documentclass[aps,prb,twocolumn,floatfix,amsmath,amssymb,
superscriptaddress]{revtex4-2}
\usepackage{amsfonts}
\usepackage{stmaryrd}
\usepackage{bbm}
\usepackage{mathrsfs}
\usepackage{tipa}
\usepackage{amssymb}
\usepackage{txfonts}
\usepackage{graphicx}
\usepackage{dcolumn}
\usepackage{epstopdf}
\usepackage[colorlinks,linkcolor=blue,urlcolor=blue,citecolor=blue]{hyperref}
\usepackage{multirow}
\usepackage{subfigure}

\begin{document}
\newcommand*{\cm}{cm$^{-1}$\,}
\newcommand*{\Tc}{T$_c$\,}
\newcommand*{\DR}{$\Delta R/R$\,}

\title{Multiple charge-density-wave gaps in LaSbTe and CeSbTe as revealed by ultrafast spectroscopy}

\author{Liye Cao}
\affiliation{Center for Advanced Quantum Studies and Department of Physics, Beijing Normal University, Beijing 100875, China}

\author{Cuiwei Zhang}
\affiliation{Beijing National Laboratory for Condensed Matter Physics and Institute of Physics, Chinese Academy of Sciences, Beijing 100190, China}

\author{Yi Yang}
\affiliation{Beijing National Laboratory for Condensed Matter Physics and Institute of Physics, Chinese Academy of Sciences, Beijing 100190, China}

\author{Lei Wang}
\affiliation{Center for Advanced Quantum Studies and Department of Physics, Beijing Normal University, Beijing 100875, China}

\author{BiXia Gao}
\affiliation{Center for Advanced Quantum Studies and Department of Physics, Beijing Normal University, Beijing 100875, China}

\author{Xinbo Wang}
\email{xinbowang@iphy.ac.cn}
\affiliation{Beijing National Laboratory for Condensed Matter Physics and Institute of Physics, Chinese Academy of Sciences, Beijing 100190, China}

\author{Youguo Shi}
\affiliation{Beijing National Laboratory for Condensed Matter Physics and Institute of Physics, Chinese Academy of Sciences, Beijing 100190, China}
\affiliation{School of Physical Sciences, University of Chinese Academy of Sciences, Beijing 100190, China}
\affiliation{Center of Materials Science and Optoelectronics Engineering, University of Chinese Academy of Sciences, Beijing  100049, China}

\author{Rongyan Chen}
\email{rychen@bnu.edu.cn}
\affiliation{Center for Advanced Quantum Studies and Department of Physics, Beijing Normal University, Beijing 100875, China}

\begin{abstract}
Utilizing ultrafast time-resolved pump-probe spectroscopy measurements, we investigate the photoinduced quasiparticle dynamics of the topological materials LaSbTe and CeSbTe. In LaSbTe, the relaxation of quasiparticles is dominated by two different mechanisms: electron-phonon coupling, and phonon-assisted electron-hole recombination. Significantly, the amplitude of photoinduced reflectivity related to the former one shows two pronounced peaks at 156 K and 263 K, indicating the occurrence of two charge density wave (CDW) phase transitions. The ultrafast responses of CeSbTe share a lot of similarities with LaSbTe, and an additional CDW phase transition at 154 K is revealed in CeSbTe. However, the slower relaxation of CeSbTe exhibits an exotic behavior that deviates from the typical phonon-assisted electron-hole recombination process, probably due to the imbalance between the electron- and hole-type carriers. Unlike LaSbTe, the relaxation times of CeSbTe vary slightly with the pump power, inferring the possible participation of 4$f$ electron in the decay process. In addition, two oscillation modes around 1 THz and 3 THz are identified in both LaSbTe and CeSbTe, which are mostly likely to be coherent phonon modes. These findings unravel the existence of multiple CDW orders in LaSbTe and CeSbTe, offering insights into the underlying physics of these systems.

\end{abstract}

\maketitle
\section{introduction}
Charge density wave (CDW) describes a spontaneous symmetry broken state, where the charge density exhibits periodically spacial modulations and is generally accompanied by a lattice distortion of the same periodicity. It is originally proposed to take place in one-dimensional electron-phonon ($e$-$ph$) coupled chains whose Fermi surfaces (FSs) only consist of two points and could be connected by a wave vector $q=2k_F$. The perfect nesting of the FSs leads to a divergence in the electronic susceptibility, and therefore the instability of the system, which is named as Peierls transition. Meanwhile, the phonon dispersion is strongly renormalized and softens to zero frequency at $q=2k_F$, which stabilizes the lattice distortion and is known as Kohn anomaly \cite{George1994Density}.
In higher dimensional materials, CDW orders are also observed frequently, even though the divergent feature of electronic susceptibility is smeared out. As a result, a number of theoretical models are developed in order to explain the exotic CDW orders in quasi-two-dimensional (2D) and even three-dimensional (3D) materials, such as $q$-dependent $e$-$ph$ coupling model \cite{Pouget_2024}, Van-hove singularity scenario and excitonic insulator model \cite{Gabovich2001}. Nevertheless, the nesting mechanism is still believed to play an important role in many 2D CDW materials which host large sections of paralleled FSs. An outstanding example is the rare-earth poly-tellurides $R$Te$_n$ ($R$ = rare earth; $n$ = 2, 2.5, 3) family, which exhibit universal CDW orders due to FS nesting \cite{PhysRevB.72.085132,PhysRevB.77.165101,PhysRevB.77.235104}.

Very recently, CDW was evidenced in some members of the quasi-2D $R$SbTe ($R$ = rare-earth elements) family, which are reported to be topological materials as well \cite{RN203,PhysRevB.103.125131,2018}. Therefore they can serve as an excellent platform to study the interplay between CDW and topological electronic structures, and potentially explore the axion electrodynamics. Particularly, the FS topology of $R$SbTe is very similar to that of $R$Te$_3$, where the former features two parallel sheets of diamond-shaped FS \cite{PhysRevB.103.125131,RN487}, and the latter consists of a double-layer FS \cite{PhysRevB.77.235104}. CDW is thereby anticipated in all of the $R$SbTe compounds due to the comparable nesting qualifications as in the $R$Te$_3$ family. Furthermore, nesting could take place along two perpendicular directions among the diamond-shaped FS, and the two parallel sheets might lead to additional CDW orders. However, CDW features are only observed in a few members of $R$SbTe, including LaSbTe, CeSbTe, GdSbTe, and HoSbTe \cite{PhysRevB.106.245145,PhysRevB.105.075111}, although their FS configurations are almost identical to each other. Additionally, multiple CDW gaps are solely demonstrated in GdSbTe \cite{PhysRevB.106.245145}. This situation inevitably raises questions on whether the existence of multiple CDW orders is a universal character among the $R$SbTe materials, like the $R$Te$_n$ family, and what the driving force of the CDW orders is.

To answer these questions, we utilize the time-resolved pump-probe (TRPP) technique to study the quasiparticle (QP) dynamics of LaSbTe and CeSbTe. LaSbTe is initially proposed to be a topological insulator by early theoretical calculations \cite{RN203}, yet a subsequent angle-resolved photoemission spectroscopy (ARPES) measurement suggests it to be a nodal-line semimetal  \cite{PhysRevB.103.125131}. Meanwhile, CeSbTe is believed to be a topological semimetal according to calculations and APRES measurements \cite{2018}. 
A comparative study on LaSbTe and CeSbTe will not only help elucidate the universal properties of $R$SbTe family, but also reveal the role played by the 4$f$ electrons. On the other hand, TRPP is a very effective tool in detecting both the single-particle excitations across small energy gaps \cite{PhysRevLett.82.4918,PhysRevLett.104.027003,PhysRevB.89.075114} and the collective excitations relevant to quantum orderings \cite{PhysRevLett.101.246402,PhysRevLett.69.1451,PhysRevLett.118.107402}. Previous infrared spectroscopy study demonstrates a CDW gap persisting up to room temperature in both LaSbTe and CeSbTe, indicating CDW phase transitions at even higher temperatures \cite{PhysRevB.106.245145}. Here, it is found that LaSbTe exhibits two more CDW transitions at 156 K and 263 K, respectively. As for CeSbTe, although we could only identify one CDW transition at 154 K, multiple CDW instabilities are believed to exist as well.

\section{experiment}

The single crystalline samples of LaSbTe and CeSbTe were synthesized by Sb-flux method \cite{PhysRevMaterials.4.094203,PhysRevB.103.125131,PhysRevB.102.155109}. TRPP measurements were performed based on the optical parameter amplifier (Orpheus F) seeded by a 50 KHz Yb:KGW amplifier (Pharos, Light Conversion) produced the laser pulses with wavelength of 800 nm and pulse duration of 60 femtoseconds. The laser pulses were split into pump and probe beams, with the former converted into 400 nm pulses through a BBO crystal. The fluence of the pump beam is about 60.27 $\mu J/cm^2$, and that of the probe beam is five times lower.

\section{results and discussions}

\subsection{Photoinduced quasiparticle dynamics of LaSbTe}

\begin{figure}[t]
  \centering
  \includegraphics[width=7.5cm]{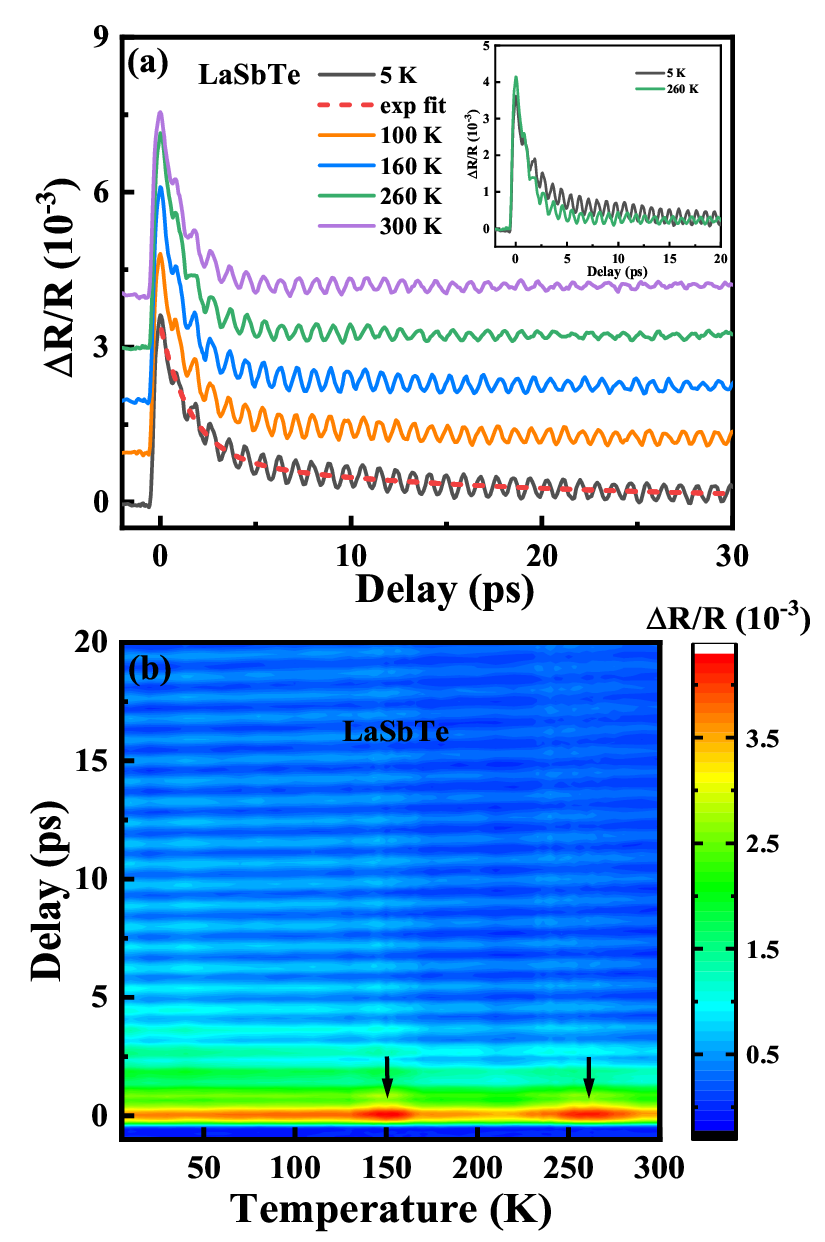}
  \caption{(a) The photoinduced reflectivity \DR of LaSbTe as a function of time delay at several selective temperatures. The red short dashed line is the single exponential fitting result of the experimental data at 5 K. The inset shows \DR at 5 K and 260 K, respectively. (b) The photoinduced reflectivity \DR as a function of time delay and temperature. The two maxima are near 150 K and 260 K, as marked by the black arrows.}
  \label{Fig:1}
\end{figure}

The photoinduced reflectivity $\Delta R/R$ at several selective temperatures of LaSbTe are shown in Fig.~\ref{Fig:1}(a). The spectra initially rise rapidly in a very short time due to temperature change induced by pump excitation, then drop back to the equilibrium state within several hundreds of picoseconds. The relaxation process can be obviously separated into two parts: the fast decay within 5 ps and the slow decay thereafter, representing different relaxation channels. Moreover, strong oscillation signals are observed in the whole temperature range. To get a better image, we plot the pseudocolor picture of \DR as a function of time delay and temperature in Fig.~\ref{Fig:1}(b). It can be seen that the amplitude of \DR exhibits two maxima near 150 K and 260 K, as indicated by the black arrows. Another characteristic is that the strength of oscillation signals is enhanced at lower temperatures and the frequency is slightly hardened at the same time.

In order to analyze the underlying mechanism in a more quantitive way, we decide to treat the relaxation process and the oscillations separately. We first decompose the picoseconds relaxation using two exponential decays,
\begin{equation}\label{Eq:1}
\Delta R/R = A_1 exp(-t/\tau_1) + A_2 exp(-t/\tau_2) + C,
\end{equation}
where $A_1$ and $A_2$ represent the amplitude of the photoinduced reflectivity change, $\tau_1$ and $\tau_2$ stand for relaxation times of different decay channels, and $C$ is used to account for the extremely long relaxation dominated by thermal diffusion. Then we analyze the residual part, which is the difference between the experimental data and the fitting results of Eq.~(\ref{Eq:1}), and derive the properties of the oscillations.

\begin{figure}[t]
  \centering
  \includegraphics[width=8.5cm]{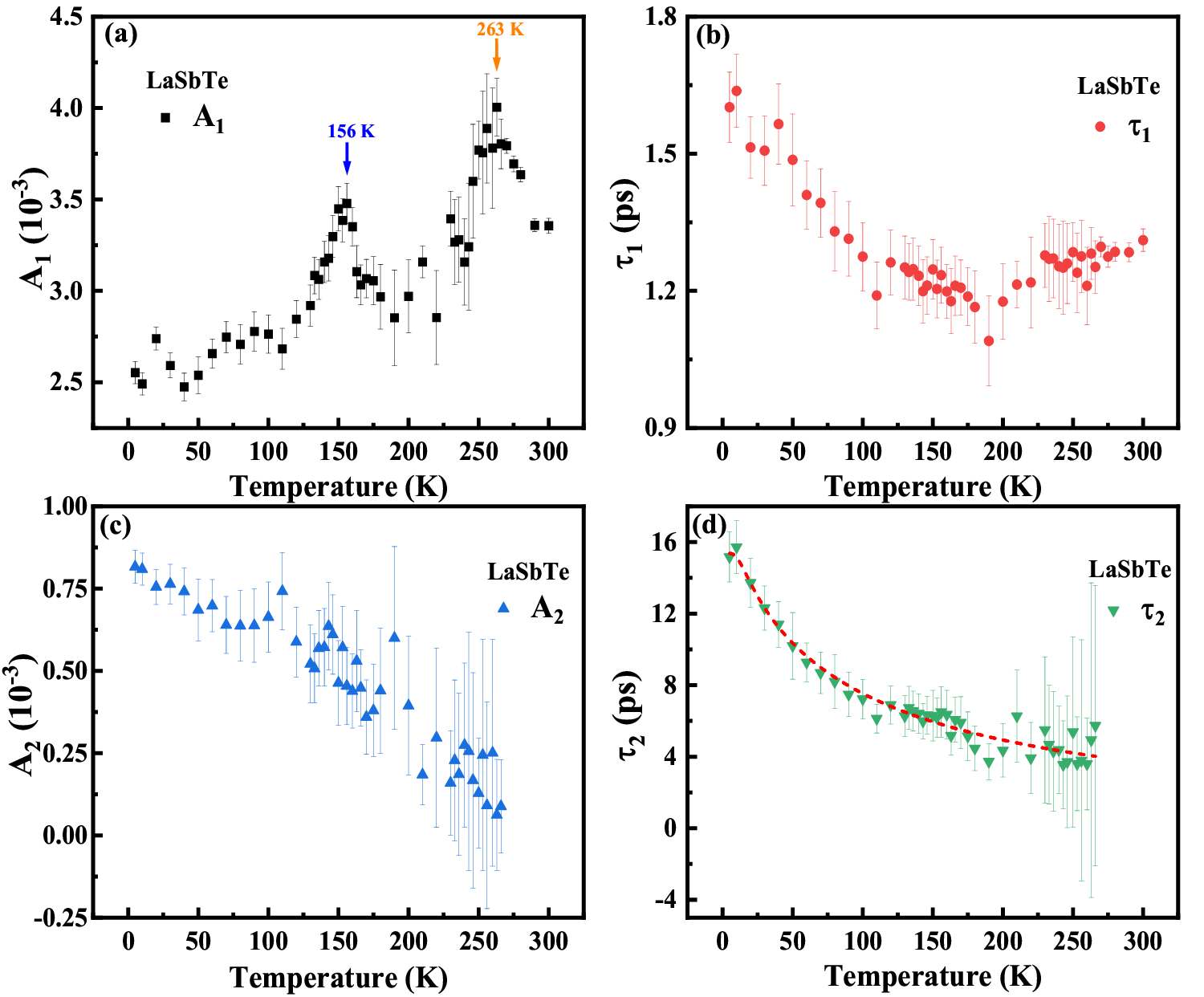}
  \caption{(a) The amplitude $A_1$ and (b) the relaxation time $\tau_1$ of the photoinduced reflectivity \DR. The blue arrow at 156 K and orange arrow at 263 K indicate the CDW phase transition temperature. (c) The amplitude $A_2$ and (d) the relaxation time $\tau_2$ of the photoinduced reflectivity \DR. The red short dashed line indicates the fitting result with the Eq.~(\ref{Eq:2}) in (d).}
  \label{Fig:2}
\end{figure}

The fitting result of \DR at 5 K based on Eq.~(\ref{Eq:1}) is shown in Fig.~\ref{Fig:1}(a) as a red short dashed line, where both the fast and slow decay parts can be well reproduced. The obtained amplitudes ($A_1$ and $A_2$) and relaxation time ($\tau_1$ and $\tau_2$) as a function of temperature are plotted in Fig.~\ref{Fig:2}. Among them, $A_1$ and $\tau_1$ correspond to the fast decay process, as shown in Figs.~\ref{Fig:2}(a) and (b). It is clearly seen that $A_1$ exhibits two pronounced peaks at around 156 K and 263 K, which is equivalent to the two maxima in Fig.~\ref{Fig:1}(b). Such kind of divergent behavior in the photoinduced quasiparticle dynamics usually points to an abrupt phase transition. However, neither theoretical calculations nor experimental measurements have demonstrated any evidence of phase transitions around these two temperatures. Although a CDW gap is revealed by our previous infrared spectroscopy, the corresponding CDW phase transition is supposed to take place well above room temperature \cite{PhysRevB.106.245145}. On the other hand, the unique double-sheet structure of the FS of LaSbTe might support the existence of multiple CDW orders, which have never been reported yet.

In order to clarify whether LaSbTe experiences more CDW phase transitions below room temperature, we resort to the phenomenal Rothwarf-Taylor (RT) model \cite{PhysRevLett.19.27}. This model was initially formulated to illustrate the ultrafast decay dynamics of photoexcited QPs in superconductors \cite{PhysRevLett.95.147002,
RTSC}, but its applicability has extended to a wide range of metallic systems with gap openings in the density of states, such as CDW and spin density wave systems \cite{PhysRevLett.101.246402,PhysRevB.89.075114,PhysRevB.101.205112,Chia-2011,LaOFeAs}. Upon photoexcitation, a significant quantity of QPs would be generated, which decay through electron-electron or $e$-$ph$ interactions to the initial equilibrium state. The emergence of an energy gap would give rise to a bottleneck phenomenon, which notably hinders the relaxation of QPs. This bottleneck effect is developed when QPs recombine across an energy gap, and generate plenty of high-energy phonons, which induce a substantial resurgence of QPs, and thus an extended relaxation time. Based on the RT model, the amplitude $A$ of photoinduced reflectivity starts to increase dramatically at the transition point, quickly reaches a maximum, and then becomes relatively stable down to the lowest temperatures. Meanwhile, the relaxation time $\tau$ exhibits a divergence at the phase transition temperature, as observed in a large number of symmetry-breaking materials \cite{RT-A,PhysRevLett.116.107001,PhysRevB.84.174412,LaOFeAs,PhysRevLett.83.800}.

At first glance, the amplitude $A_1$ of LaSbTe only shows two peak structures and no divergences could be identified in the temperature dependent relaxation time $\tau_1$, which seems to be incompatible with the RT model. However, these characters are reminiscent of the pump-probe spectroscopy of $R$Te$_3$ system ($R$ = Tb, Dy, and Ho) \cite{PhysRevB.89.075114,PhysRevLett.101.246402} at the low temperature region. These materials are known to undergo multiple CDW phase transitions. Across the higher temperature one ($T_{C1}$), there is a sharp increase in amplitude along with a divergence in the relaxation time, agreeing well with the RT model. Across the lower temperature one ($T_{C2}$), however, the increase of the amplitude is much less pronounced and a peak feature is observed for TbTe$_3$, just like the results of LaSbTe observed here. Moreover, the relaxation time is hardly changed across $T_{C2}$ for all the three $R$Te$_3$ compounds, analogy to the behavior of $\tau_1$ in Fig.~\ref{Fig:2}(b). Considering that there is already a large CDW gap at room temperature in LaSbTe \cite{PhysRevB.106.245145}, we believe the two peaks at 156 K and 263 K indicate two additional CDW phase transitions. It is very likely that these multiple CDW orders originate from the nesting between the same part of FS and thus compete with each other, just like the situation of $R$Te$_3$ \cite{PhysRevB.77.235104,PhysRevB.90.085105}. Most parts of the FS are gaped by the highest temperature one, leaving only a tiny part for the second and third transition, which is therefore much weaker. As a result, typical transient characters defined by the RT model are too weak to be observed.

Different from the transient dynamics of $R$Te$_3$, where the photoinduced quasiparticles decay through one picosecond channel, there is an additional slower one in LaSbTe. As shown in Fig.~\ref{Fig:2}(c), the amplitude $A_2$ of this slow decay process emerges at about 270 K and gradually increases as the temperature decreases. The relaxation time $\tau_2$ also increases upon cooling, as plotted in Fig.~\ref{Fig:2}(d). Considering LaSbTe is a nodal-line semimetal with both electron and hole pockets \cite{PhysRevB.103.125131,RN487,PhysRevB.96.245138}, we propose that the relaxation time $\tau_2$ may be associated with the so called phonon-assisted electron-hole ($e$-$h$) recombination between the conduction and valence bands, which is widely observed in a number of semimetal materials \cite{PdTe2,PhysRevB.92.161104,PhysRevB.103.115108,PhysRevB.106.104308}. This process generally emerges afterwards the $e$-$ph$ thermalization, and lasts for about several or tens of ps, in perfect agreement with our results.
Since the electron and hole Fermi pockets are usually distinctly separated in momentum space, the $e$-$h$ recombination is achieved through the assistance of phonons, in order to meet the momentum conservation requirements. The T-dependence of the phonon-assisted $e$-$h$ recombination time can be quantitatively described by \cite{PhysRev.175.823}
\begin{equation}\label{Eq:2}
\frac{1}{\tau} = \Phi \frac{\frac{\hbar\omega}{2k_B T}}{\sinh^2(\frac{\hbar\omega}{2k_B T})} + \frac{1}{\tau_0},
\end{equation}
where $\omega$ is the frequency of the phonon mode assisting $e$-$h$ recombination, $\tau_0$ represents the temperature-independent recombination time, and $\Phi$ is the fitting parameter. The red short dashed line in Fig.~\ref{Fig:2}(d) represents the least-squares fit to the data, yielding $\tau_0$ = 15.4 ps. It suggests a good match with Eq.~(\ref{Eq:2}) in whole measured temperature ranges.

\begin{figure}[t]
  \centering
  \includegraphics[width=8.5cm]{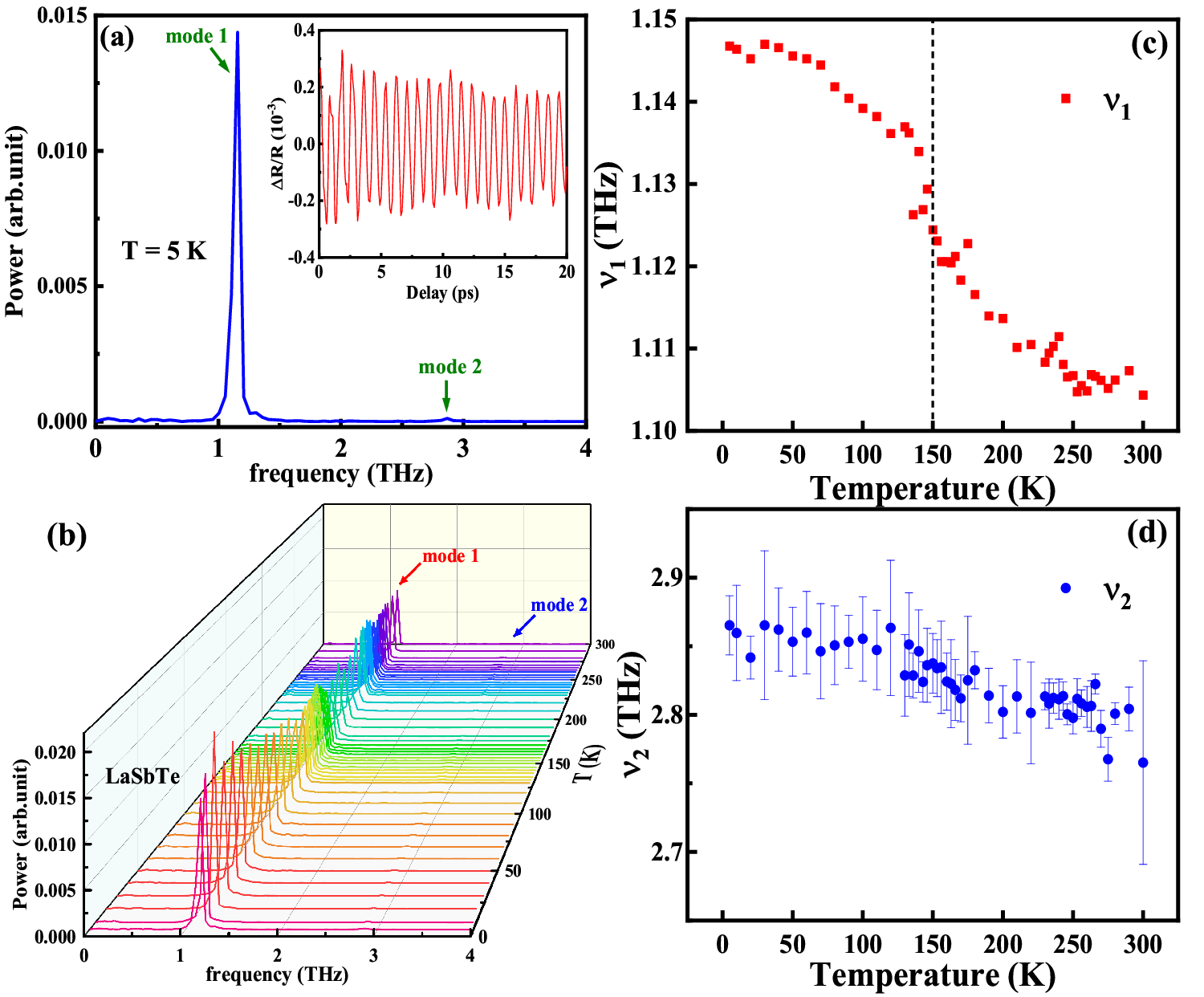}
  \caption{(a) The fast Fourier transformation of the oscillation part of the photoinduced reflectivity \DR at 5 K. The inset displays the corresponding residual oscillation part of \DR. (b) The fast Fourier transformation results of the oscillation part of \DR as a function of frequency and temperature. The two peaks are indicated by the red and blue arrows, respectively. The temperature-dependent frequencies of the two modes are depicted in (c) for $\nu_1$ (red squares)(blue circles) and (d) for $\nu_2$ (blue circles). The black short dashed line indicates the location of the bend.}
  \label{Fig:3}
\end{figure}

Notably, significant oscillatory signals are observed at all measured temperatures, the lifetime and periods of which change with temperature variation. In order to analyze the oscillatory component quantitatively, we first subtract the exponential fitting results from the experimental data in order to derive the residual oscillation signals. We show the residual oscillation signal at 5 K in the inset of Fig.~\ref{Fig:3}(a). The frequencies of the oscillations could be obtained by performing fast Fourier transformations (FFT) on the residual signals. As displayed in the main panel in Fig.~\ref{Fig:3}(a), two oscillation modes could be resolved, centering at 1.15 and 2.87 THz, respectively. We mark these two peaks as ``mode 1'' and ``mode 2'' for convenience. The FFT results of the oscillation part from 5 K to 300 K are plotted in Fig.~\ref{Fig:3}(b), which demonstrates the existence of ``mode 1'' and ``mode 2'' at all measured temperatures.

To track the evolvement of the two modes with temperature, we have extracted their frequencies and plotted the Gauss fitting results of frequency peaks in Figs.~\ref{Fig:3}(c) and (d) as red squares ($\nu_1$) and blue circles ($\nu_2$). It is clearly shown that $\nu_1$ and $\nu_2$ both soften extremely slightly with temperature increasing, which is inconsistent with the behavior of an amplitude mode. Therefore, they are most likely to be coherent phonons, and the slight softening is probably due to thermal effect. However, since there is supposed to be another CDW phase transition above room temperature, we could not completely rule out the possibility that one of the two modes is actually an amplitude mode. Particularly, $\nu_1$ exhibits an obvious drop near 150 K, which is close to one of the CDW phase transition temperatures, indicated by a black short dashed line in Fig.~\ref{Fig:3}(c). We propose this behavior is caused by the mild lattice modulation related to the CDW phase transition at 156 K.

\begin{figure}[t]
  \centering
  \includegraphics[width=8.5cm]{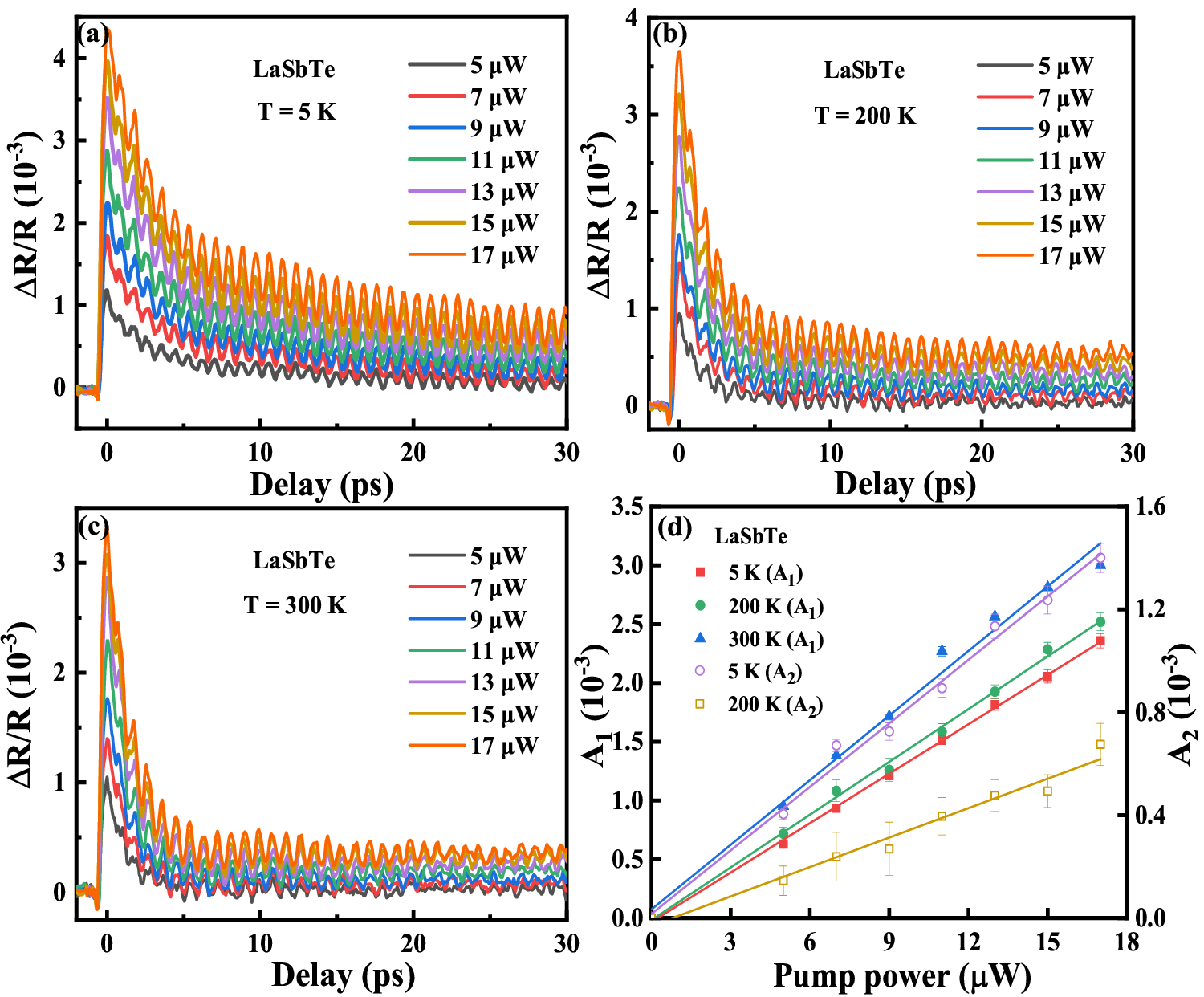}
  \caption{The photoinduced reflectivity under different pump power at (a) 5 K, (b) 200 K, and (c) 300 K. It's noted that the illuminating spot remains unchanged during the measurements and 5 $\mu W$ give rise to a fluence of about 38 $\mu J/cm^2$. (d) The pump power dependence of $A_1$ at 5 K, 200 K, and 300 K, as well as $A_2$ at 5 K and 200 K. The solid lines are the results of linear fitting of experimental data.}
  \label{Fig:4}
\end{figure}

In order to confirm our assignment of the relaxation process, we also investigate the QP dynamics at different fluence. Since there is already a CDW gap at room temperature, and the second and third CDW phase transitions take place at 263 K and 156 K, we choose 5 K, 200 K, and 300 K to represent the three different states. The respective results obtained at these three temperatures are displayed in Figs.~\ref{Fig:4}(a)-(c). By fitting them according to Eq.~(\ref{Eq:1}), we can get the fluence dependent behavior of the amplitudes and relaxation times of the two channels. We find that the relaxation times remain constant against fluence changing all along. Meanwhile, the amplitudes $A_1$ and $A_2$ increase linearly with fluence at all the measured temperatures, as depicted in Fig.~\ref{Fig:4}(d). This phenomenon is frequently observed in studies on CDW transition \cite{PhysRevLett.118.107402} and $e$-$h$ recombination \cite{PhysRevB.87.075429} dynamics. This observation indicates that the number of excited QPs is proportional to photo-induced reflectivity, and does not reach saturation for both decay channels. In other words, neither the energy gap of the CDW ordering nor the phonon-assisted $e$-$h$ recombination is quenched by the laser excitation up to our highest fluence. Before moving to the next subject, we want to remark on the small dip emerging prior to the rising edge, which recovers within 0.2 ps. This phenomenon is only observed occasionally during our measurements, suggesting that it is most likely an artifact.

\subsection{Photoinduced quasiparticle dynamics of CeSbTe}

\begin{figure}[t]
  \centering
  \includegraphics[width=7.5cm]{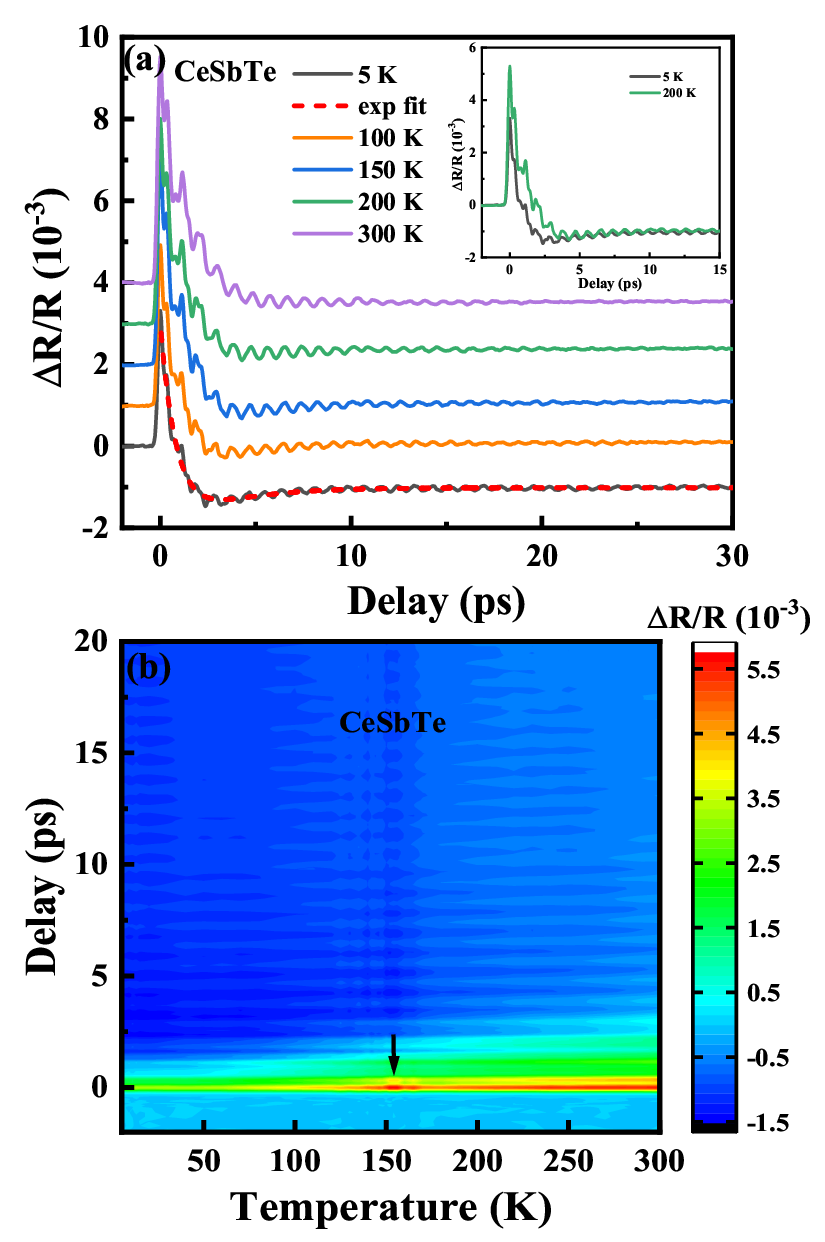}
  \caption{(a) The photoinduced reflectivity \DR of CeSbTe as a function of time delay at several selective temperatures. The red dashed line is the single exponential fitting result of the experimental data at 5 K. The inset shows \DR at 5 K and 200 K, respectively. (b) The photoinduced reflectivity \DR as a function of time delay and temperature. The maximum occurs near 150 K, as marked by the black arrow.}
  \label{Fig:5}
\end{figure}

Figure~\ref{Fig:5}(a) displays the photoinduced change of reflectivity \DR of CeSbTe at several selective temperatures. \DR of each spectrum undergoes a dynamic process roughly similar to LaSbTe. We observe three distinct decay processes: a fast decay process with a positive amplitude value, a slow decay process with a negative amplitude value, and a much longer relaxation associated with the heat diffusion process. Furthermore, pronounced oscillation signals are evident across the entire temperature range. To provide a more vivid depiction, the raw experimental data for all measured temperatures are plotted in a pseudocolor image, as shown in Fig.~\ref{Fig:5}(b). A maximum of \DR near 150 K is observed, marked by a black arrow.

\begin{figure}[t]
  \centering
  \includegraphics[width=8.5cm]{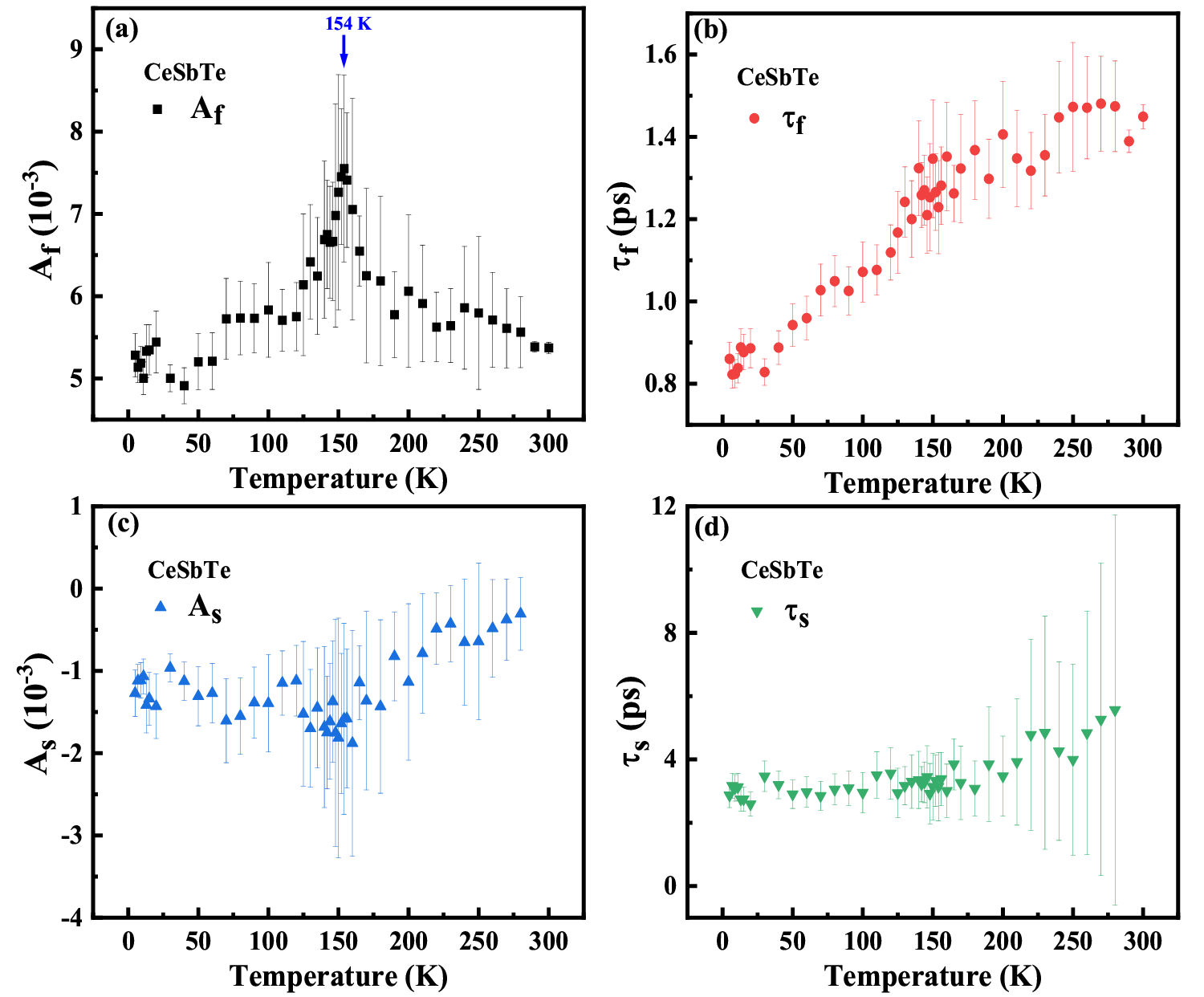}
  \caption{(a) The amplitude $A_f$ and (b) the relaxation time $\tau_f$ of the photoinduced reflectivity \DR. The blue arrow at 154 K masks a CDW phase transition temperature of CeSbTe. (c) The amplitude $A_s$ and (d) the relaxation time $\tau_s$ of the photoinduced reflectivity \DR.}
  \label{Fig:6}
\end{figure}

To quantitatively analyze the physical mechanisms, we use the same strategy as we deal with LaSbTe. Firstly, we fit the data with \DR = $A_f exp(-t/\tau_f) + A_s exp(-t/\tau_s)$, where $A_f$ and $\tau_f$ denote the amplitude and relaxation time of the fast decay process, while $A_s$ and $\tau_s$ present those of the slow decay process. The fitting result of 5 K is shown in Fig.~\ref{Fig:5}(a), as indicated by a red short dashed line. We plot the amplitude $A_f$ in Fig.~\ref{Fig:6}(a), which exhibits an obvious emergence of a peak-like feature with its center at 154 K, as pointed out by a blue arrow. It is in accordance with the maximum shown in Fig.~\ref{Fig:5}(b). In the meantime, the relaxation time $\tau_f$ gradually increases upon warming, without any noticeable anomalies. All these behaviors are highly analogous to that of LaSbTe, except that the temperature dependent amplitude $A_f$ only shows one peak instead of two. Based on the above discussions of LaSbTe and the similarity between LaSbTe and CeSbTe, we believe it is safe to conclude that the peak feature at 154 K shown in Fig.~\ref{Fig:6}(a) is indicative of a CDW transition in CeSbTe. Additionally, as another CDW transition is suggested to occur at a temperature exceeding 300 K in our previous infrared spectroscopy measurement \cite{PhysRevB.106.245145}, at least two CDW orderings are expected to exist in CeSbTe.

In contrast to the fast relaxation process linked with QPs crossing the energy gap, the slow relaxation process of \DR in CeSbTe represents another decay channel. We plot the amplitude $A_s$ and relaxation time $\tau_s$ in Fig.~\ref{Fig:6}(c) and (d), respectively. Although this slow decay emerges slightly below room temperature and gets more pronounced with temperature decreasing, similar to LaSbTe, the amplitude $A_s$ exhibits a negative value, which is in sharp contrast to $A_2$ of LaSbTe. Moreover, the relaxation time $\tau_s$ is hardly changed with temperature variation, which can not be well fitted by Eq.~(\ref{Eq:2}). Consequently, phonon-assisted $e$-$h$ recombination no longer seems suitable to explain the origin of this slow decay in CeSbTe. Although $e$-$h$ recombination can also occur through the emission of photons or via the three-particle Auger process \cite{10.1063/1.367411,PhysRevB.46.13194}, these processes usually take place on much longer time scales \cite{10.1063/1.367411}, which does not align with the case of CeSbTe.

\begin{figure}[t]
  \centering
  \includegraphics[width=8.5cm]{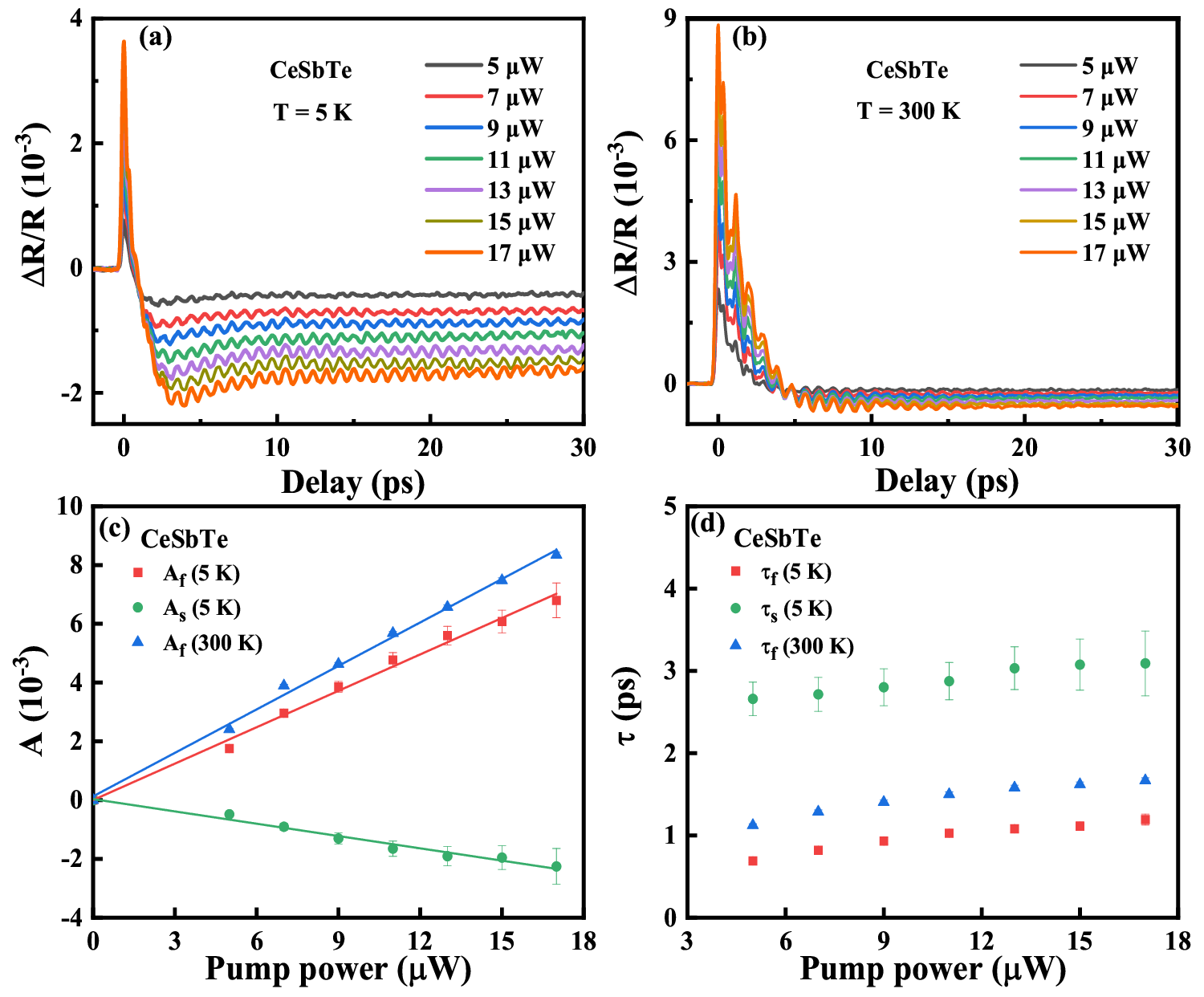}
  \caption{The photoinduced reflectivity under different pump power at (a) 5 K and (b) 300 K. It's noted that the illuminating spot remains unchanged during the measurements and 5 $\mu W$ give rise to a fluence of about 38 $\mu J/cm^2$. (c) The pump power dependence of $A_f$ at 5 K and 300 K, as well as $A_s$ at 5 K. The solid lines are the results of linear fitting of experimental data. (d) The pump power dependence of $\tau_f$ at 5 K and 300 K, as well as $\tau_s$ at 5 K.}
  \label{Fig:7}
\end{figure}

To gain further insight into the ultrafast dynamic responses of CeSbTe, we also measure the photoinduced reflectivity \DR change at different pump fluence. Given the presence of a CDW gap at room temperature and a second CDW phase transition occurring at 154 K, we select temperatures of 5 K and 300 K to represent the two distinct states, as displayed in Figs.~\ref{Fig:7}(a) and (b). The obtained amplitudes and relaxation times are shown in Figs.~\ref{Fig:7}(c) and (d), respectively. Although there are two decay channels at 5 K and only one at 300 K, all the amplitudes show a proportional relationship with the pump power, indicating a linear increase in the number of excited QPs. This suggests that even at the highest laser pulse used, the decay dynamics remain robust, consistent with observations in LaSbTe. On the other hand, the relaxation time $\tau_f$ and $\tau_s$ both exhibit a slight enhancement with increasing fluence, which differs from that observed in LaSbTe.

To clarify the underlying mechanism of the exotic quasiparticle dynamics of CeSbTe, we have compared its physical properties with those of LaSbTe, and found that they have quite different electric transport properties. The temperature-dependent resistivity of LaSbTe exhibits metallic characteristics \cite{PhysRevB.96.245138}, whereas CeSbTe presents a weak insulating behavior \cite{Lv_2019}. Furthermore, Hall measurements reveal that the electron and hole densities are nearly equivalent in LaSbTe, indicating a near-perfect compensation of $e$-$h$ carriers \cite{PhysRevB.96.245138}. In contrast, CeSbTe is dominated by hole-type carriers, especially at low temperatures \cite{Lv_2019}. This imbalance in carrier types might affect the phonon-assisted $e$-$h$ recombination process, and thus lead to the unique behavior of $\tau_s$. Additionally, the magnetic susceptibility of CeSbTe is much larger than that of LaSbTe even in the paramagnetic state, due to the existence of 4$f$ electrons in Ce. We speculate that the local moment of the 4$f$ electrons might somehow participate in the relaxation process of excited quasiparticles, and affect the fluence dependent behavior of the relaxation times. Further researches are required to thoroughly elucidate the underlying mechanisms of this process.

\begin{figure}[t]
  \centering
  \includegraphics[width=8.5cm]{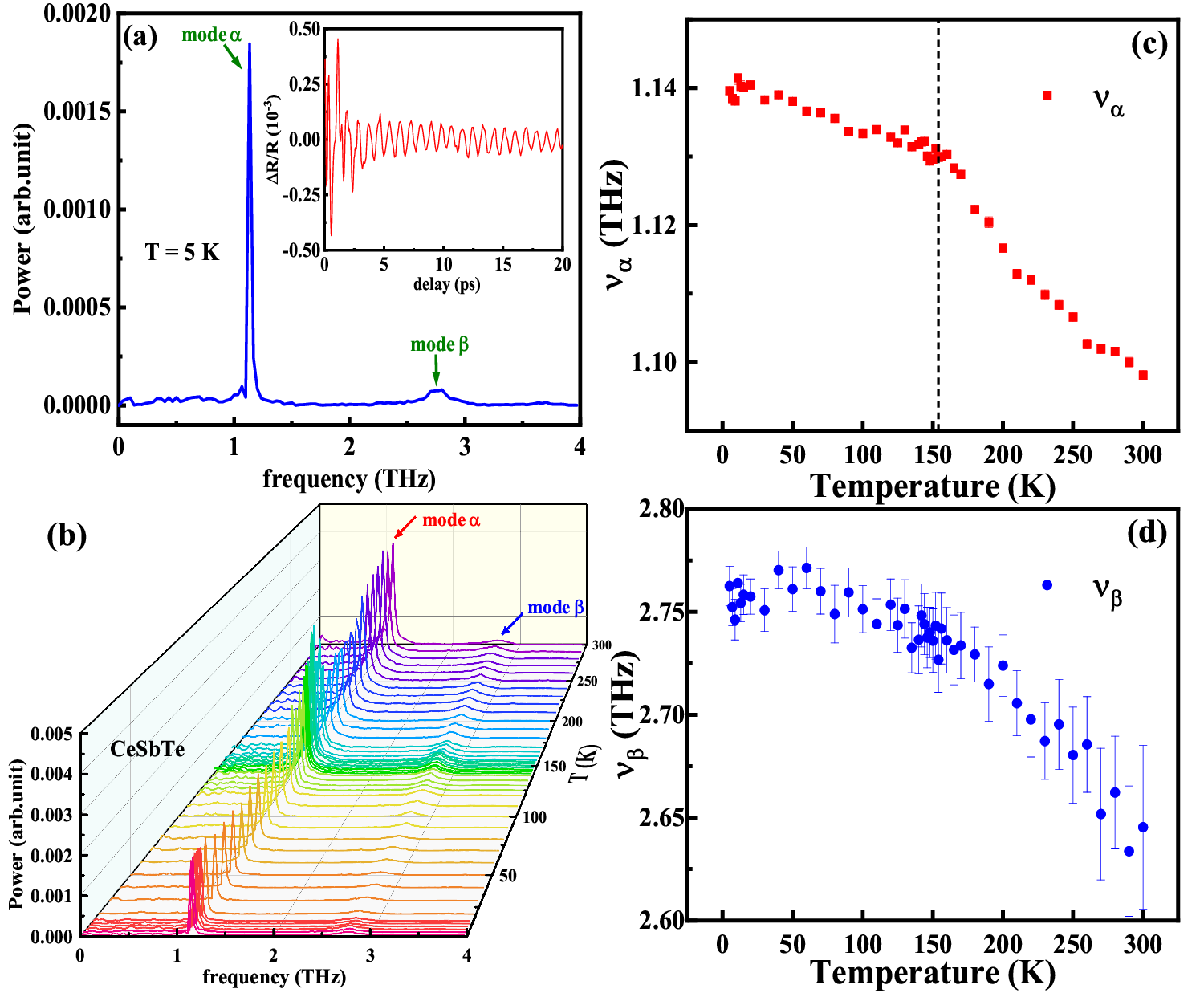}
  \caption{(a) The fast Fourier transformation of the oscillation part of the photoinduced reflectivity \DR at 5 K. The inset displays the corresponding residual oscillation part of \DR. (b) The fast Fourier transformation results of the oscillation part of \DR as a function of frequency and temperature. The two peaks are indicated by the red and blue arrows, respectively. The temperature-dependent frequencies of the two modes are depicted in (c) for $\nu_\alpha$ (red squares) and (d) for $\nu_\beta$ (blue circles). The black short dashed line indicates the location of the bend.}
  \label{Fig:8}
\end{figure}

Pronounced oscillatory signals are observed in \DR of CeSbTe as well. Employing a quantitative analysis akin to that of LaSbTe, we illustrate the residual oscillation signal at 5 K in the inset of Fig.~\ref{Fig:8}(a), and the main panel displays the corresponding FFT results in 0 - 4 THz. Two distinct peaks located at about 1.14 THz and 2.76 THz could be resolved, which are marked as ``mode $\alpha$'' and ``mode $\beta$''. The FFT results in the whole temperature range are presented in Fig.~\ref{Fig:8}(b). The frequencies of the two modes as a function of temperature are plotted in Figs.~\ref{Fig:8}(c) and (d), labeled with red squares ($\nu_\alpha$) and blue circles ($\nu_\beta$). Similar to LaSbTe, these two modes soften very slightly upon warming, indicating they are mostly likely to be coherent phonons. Moreover, ``mode $\alpha$'' undergoes a noticeable drop near 150 K as marked by a black short dashed line in Fig.~\ref{Fig:8}(c), which is possibly related to the CDW phase transition identified by the amplitude $A_f$.

\section{conclusion}

In conclusion, we have conducted ultrafast pump-probe measurements on the topological materials LaSbTe and CeSbTe. The photoinduced QPs in LaSbTe primarily relax through two decay processes. The amplitude of the first relaxation process exhibits two distinct peak features at 156 K and 263 K, which suggests two CDW phase transitions that have never been reported before. Meanwhile, the second decay is believed to be associated with a phonon-assisted $e$-$h$ recombination process. In CeSbTe, a similar fast relaxation process akin to that in LaSbTe is observed, indicating an additional CDW phase transition at 154 K. However, the second relaxation process in CeSbTe exhibits characteristics different from LaSbTe, potentially related to the imbalance between electron and hole carriers. Moreover, the relaxation times of CeSbTe increased slightly with increasing pump power, possibly associated with the local moments of the 4$f^1$ electron in Ce. Besides, two oscillation modes around 1 THz and 3 THz are identified in both compounds. Our results provide important information towards a profound understanding of the $R$SbTe system and inspire future researches on the subject.

\begin{center}
\small{\textbf{ACKNOWLEDGMENTS}}
\end{center}

This work was supported by the National Key Projects for Research and Development of China (Grant No. 2021YFA1400400), the National Natural Science Foundation of China (Grants No. 12074042, No. U2032204, No. 11974414, No. U22A6005, and No. 11704033), the Strategic Priority Research Program of the Chinese Academy of Sciences (Grant No. XDB33030000), 
and the Fundamental Research Funds for the Central Universities (Grant No. 2243300003). This work was supported by the Synergetic Extreme Condition User Facility (SECUF).

\bibliographystyle{apsrev4-2}
  \bibliography{ReferenceLnSbTe}

\begin{thebibliography}{41}%
\makeatletter
\providecommand \@ifxundefined [1]{%
 \@ifx{#1\undefined}
}%
\providecommand \@ifnum [1]{%
 \ifnum #1\expandafter \@firstoftwo
 \else \expandafter \@secondoftwo
 \fi
}%
\providecommand \@ifx [1]{%
 \ifx #1\expandafter \@firstoftwo
 \else \expandafter \@secondoftwo
 \fi
}%
\providecommand \natexlab [1]{#1}%
\providecommand \enquote  [1]{``#1''}%
\providecommand \bibnamefont  [1]{#1}%
\providecommand \bibfnamefont [1]{#1}%
\providecommand \citenamefont [1]{#1}%
\providecommand \href@noop [0]{\@secondoftwo}%
\providecommand \href [0]{\begingroup \@sanitize@url \@href}%
\providecommand \@href[1]{\@@startlink{#1}\@@href}%
\providecommand \@@href[1]{\endgroup#1\@@endlink}%
\providecommand \@sanitize@url [0]{\catcode `\\12\catcode `\$12\catcode
  `\&12\catcode `\#12\catcode `\^12\catcode `\_12\catcode `\%12\relax}%
\providecommand \@@startlink[1]{}%
\providecommand \@@endlink[0]{}%
\providecommand \url  [0]{\begingroup\@sanitize@url \@url }%
\providecommand \@url [1]{\endgroup\@href {#1}{\urlprefix }}%
\providecommand \urlprefix  [0]{URL }%
\providecommand \Eprint [0]{\href }%
\providecommand \doibase [0]{https://doi.org/}%
\providecommand \selectlanguage [0]{\@gobble}%
\providecommand \bibinfo  [0]{\@secondoftwo}%
\providecommand \bibfield  [0]{\@secondoftwo}%
\providecommand \translation [1]{[#1]}%
\providecommand \BibitemOpen [0]{}%
\providecommand \bibitemStop [0]{}%
\providecommand \bibitemNoStop [0]{.\EOS\space}%
\providecommand \EOS [0]{\spacefactor3000\relax}%
\providecommand \BibitemShut  [1]{\csname bibitem#1\endcsname}%
\let\auto@bib@innerbib\@empty
\bibitem [{\citenamefont {Gr{\"u}ner}(1994)}]{George1994Density}%
  \BibitemOpen
  \bibfield  {author} {\bibinfo {author} {\bibfnamefont {G.}~\bibnamefont
  {Gr{\"u}ner}},\ }\href@noop {} {\emph {\bibinfo {title} {Density waves in
  solids}}}\ (\bibinfo  {publisher} {Perseus, Cambridge, MA},\ \bibinfo {year}
  {1994})\BibitemShut {NoStop}%
\bibitem [{\citenamefont {Pouget}\ and\ \citenamefont
  {Canadell}(2024)}]{Pouget_2024}%
  \BibitemOpen
  \bibfield  {author} {\bibinfo {author} {\bibfnamefont {J.-P.}\ \bibnamefont
  {Pouget}}\ and\ \bibinfo {author} {\bibfnamefont {E.}~\bibnamefont
  {Canadell}},\ }\href {https://doi.org/10.1088/1361-6633/ad124f} {\bibfield
  {journal} {\bibinfo  {journal} {Reports on Progress in Physics}\ }\textbf
  {\bibinfo {volume} {87}},\ \bibinfo {pages} {026501} (\bibinfo {year}
  {2024})}\BibitemShut {NoStop}%
\bibitem [{\citenamefont {Gabovich}\ \emph {et~al.}(2001)\citenamefont
  {Gabovich}, \citenamefont {Voitenko}, \citenamefont {Annett},\ and\
  \citenamefont {Ausloos}}]{Gabovich2001}%
  \BibitemOpen
  \bibfield  {author} {\bibinfo {author} {\bibfnamefont {A.~M.}\ \bibnamefont
  {Gabovich}}, \bibinfo {author} {\bibfnamefont {A.~I.}\ \bibnamefont
  {Voitenko}}, \bibinfo {author} {\bibfnamefont {J.~F.}\ \bibnamefont
  {Annett}},\ and\ \bibinfo {author} {\bibfnamefont {M.}~\bibnamefont
  {Ausloos}},\ }\href {https://doi.org/10.1088/0953-2048/14/4/201} {\bibfield
  {journal} {\bibinfo  {journal} {Superconductor Science and Technology}\
  }\textbf {\bibinfo {volume} {14}},\ \bibinfo {pages} {R1} (\bibinfo {year}
  {2001})}\BibitemShut {NoStop}%
\bibitem [{\citenamefont {Shin}\ \emph {et~al.}(2005)\citenamefont {Shin},
  \citenamefont {Brouet}, \citenamefont {Ru}, \citenamefont {Shen},\ and\
  \citenamefont {Fisher}}]{PhysRevB.72.085132}%
  \BibitemOpen
  \bibfield  {author} {\bibinfo {author} {\bibfnamefont {K.~Y.}\ \bibnamefont
  {Shin}}, \bibinfo {author} {\bibfnamefont {V.}~\bibnamefont {Brouet}},
  \bibinfo {author} {\bibfnamefont {N.}~\bibnamefont {Ru}}, \bibinfo {author}
  {\bibfnamefont {Z.~X.}\ \bibnamefont {Shen}},\ and\ \bibinfo {author}
  {\bibfnamefont {I.~R.}\ \bibnamefont {Fisher}},\ }\href
  {https://doi.org/10.1103/PhysRevB.72.085132} {\bibfield  {journal} {\bibinfo
  {journal} {Phys. Rev. B}\ }\textbf {\bibinfo {volume} {72}},\ \bibinfo
  {pages} {085132} (\bibinfo {year} {2005})}\BibitemShut {NoStop}%
\bibitem [{\citenamefont {Shin}\ \emph {et~al.}(2008)\citenamefont {Shin},
  \citenamefont {Laverock}, \citenamefont {Wu}, \citenamefont {Condron},
  \citenamefont {Toney}, \citenamefont {Dugdale}, \citenamefont {Kramer},\ and\
  \citenamefont {Fisher}}]{PhysRevB.77.165101}%
  \BibitemOpen
  \bibfield  {author} {\bibinfo {author} {\bibfnamefont {K.~Y.}\ \bibnamefont
  {Shin}}, \bibinfo {author} {\bibfnamefont {J.}~\bibnamefont {Laverock}},
  \bibinfo {author} {\bibfnamefont {Y.~Q.}\ \bibnamefont {Wu}}, \bibinfo
  {author} {\bibfnamefont {C.~L.}\ \bibnamefont {Condron}}, \bibinfo {author}
  {\bibfnamefont {M.~F.}\ \bibnamefont {Toney}}, \bibinfo {author}
  {\bibfnamefont {S.~B.}\ \bibnamefont {Dugdale}}, \bibinfo {author}
  {\bibfnamefont {M.~J.}\ \bibnamefont {Kramer}},\ and\ \bibinfo {author}
  {\bibfnamefont {I.~R.}\ \bibnamefont {Fisher}},\ }\href
  {https://doi.org/10.1103/PhysRevB.77.165101} {\bibfield  {journal} {\bibinfo
  {journal} {Phys. Rev. B}\ }\textbf {\bibinfo {volume} {77}},\ \bibinfo
  {pages} {165101} (\bibinfo {year} {2008})}\BibitemShut {NoStop}%
\bibitem [{\citenamefont {Brouet}\ \emph {et~al.}(2008)\citenamefont {Brouet},
  \citenamefont {Yang}, \citenamefont {Zhou}, \citenamefont {Hussain},
  \citenamefont {Moore}, \citenamefont {He}, \citenamefont {Lu}, \citenamefont
  {Shen}, \citenamefont {Laverock}, \citenamefont {Dugdale}, \citenamefont
  {Ru},\ and\ \citenamefont {Fisher}}]{PhysRevB.77.235104}%
  \BibitemOpen
  \bibfield  {author} {\bibinfo {author} {\bibfnamefont {V.}~\bibnamefont
  {Brouet}}, \bibinfo {author} {\bibfnamefont {W.~L.}\ \bibnamefont {Yang}},
  \bibinfo {author} {\bibfnamefont {X.~J.}\ \bibnamefont {Zhou}}, \bibinfo
  {author} {\bibfnamefont {Z.}~\bibnamefont {Hussain}}, \bibinfo {author}
  {\bibfnamefont {R.~G.}\ \bibnamefont {Moore}}, \bibinfo {author}
  {\bibfnamefont {R.}~\bibnamefont {He}}, \bibinfo {author} {\bibfnamefont
  {D.~H.}\ \bibnamefont {Lu}}, \bibinfo {author} {\bibfnamefont {Z.~X.}\
  \bibnamefont {Shen}}, \bibinfo {author} {\bibfnamefont {J.}~\bibnamefont
  {Laverock}}, \bibinfo {author} {\bibfnamefont {S.~B.}\ \bibnamefont
  {Dugdale}}, \bibinfo {author} {\bibfnamefont {N.}~\bibnamefont {Ru}},\ and\
  \bibinfo {author} {\bibfnamefont {I.~R.}\ \bibnamefont {Fisher}},\ }\href
  {https://doi.org/10.1103/PhysRevB.77.235104} {\bibfield  {journal} {\bibinfo
  {journal} {Phys. Rev. B}\ }\textbf {\bibinfo {volume} {77}},\ \bibinfo
  {pages} {235104} (\bibinfo {year} {2008})}\BibitemShut {NoStop}%
\bibitem [{\citenamefont {Qian}\ \emph {et~al.}(2020)\citenamefont {Qian},
  \citenamefont {Tan}, \citenamefont {Zhang}, \citenamefont {Gao},
  \citenamefont {Wang}, \citenamefont {Fang}, \citenamefont {Fang},\ and\
  \citenamefont {Weng}}]{RN203}%
  \BibitemOpen
  \bibfield  {author} {\bibinfo {author} {\bibfnamefont {Y.}~\bibnamefont
  {Qian}}, \bibinfo {author} {\bibfnamefont {Z.}~\bibnamefont {Tan}}, \bibinfo
  {author} {\bibfnamefont {T.}~\bibnamefont {Zhang}}, \bibinfo {author}
  {\bibfnamefont {J.}~\bibnamefont {Gao}}, \bibinfo {author} {\bibfnamefont
  {Z.}~\bibnamefont {Wang}}, \bibinfo {author} {\bibfnamefont {Z.}~\bibnamefont
  {Fang}}, \bibinfo {author} {\bibfnamefont {C.}~\bibnamefont {Fang}},\ and\
  \bibinfo {author} {\bibfnamefont {H.}~\bibnamefont {Weng}},\ }\href
  {https://doi.org/10.1007/s11433-019-1515-4} {\bibfield  {journal} {\bibinfo
  {journal} {Science China Physics, Mechanics \& Astronomy}\ }\textbf {\bibinfo
  {volume} {63}},\ \bibinfo {pages} {107011} (\bibinfo {year}
  {2020})}\BibitemShut {NoStop}%
\bibitem [{\citenamefont {Wang}\ \emph {et~al.}(2021)\citenamefont {Wang},
  \citenamefont {Qian}, \citenamefont {Yang}, \citenamefont {Chen},
  \citenamefont {Li}, \citenamefont {Tan}, \citenamefont {Cai}, \citenamefont
  {Zhao}, \citenamefont {Gao}, \citenamefont {Feng}, \citenamefont {Kumar},
  \citenamefont {Schwier}, \citenamefont {Zhao}, \citenamefont {Weng},
  \citenamefont {Shi}, \citenamefont {Wang}, \citenamefont {Song},
  \citenamefont {Huang}, \citenamefont {Shimada}, \citenamefont {Xu},
  \citenamefont {Zhou},\ and\ \citenamefont {Liu}}]{PhysRevB.103.125131}%
  \BibitemOpen
  \bibfield  {author} {\bibinfo {author} {\bibfnamefont {Y.}~\bibnamefont
  {Wang}}, \bibinfo {author} {\bibfnamefont {Y.}~\bibnamefont {Qian}}, \bibinfo
  {author} {\bibfnamefont {M.}~\bibnamefont {Yang}}, \bibinfo {author}
  {\bibfnamefont {H.}~\bibnamefont {Chen}}, \bibinfo {author} {\bibfnamefont
  {C.}~\bibnamefont {Li}}, \bibinfo {author} {\bibfnamefont {Z.}~\bibnamefont
  {Tan}}, \bibinfo {author} {\bibfnamefont {Y.}~\bibnamefont {Cai}}, \bibinfo
  {author} {\bibfnamefont {W.}~\bibnamefont {Zhao}}, \bibinfo {author}
  {\bibfnamefont {S.}~\bibnamefont {Gao}}, \bibinfo {author} {\bibfnamefont
  {Y.}~\bibnamefont {Feng}}, \bibinfo {author} {\bibfnamefont {S.}~\bibnamefont
  {Kumar}}, \bibinfo {author} {\bibfnamefont {E.~F.}\ \bibnamefont {Schwier}},
  \bibinfo {author} {\bibfnamefont {L.}~\bibnamefont {Zhao}}, \bibinfo {author}
  {\bibfnamefont {H.}~\bibnamefont {Weng}}, \bibinfo {author} {\bibfnamefont
  {Y.}~\bibnamefont {Shi}}, \bibinfo {author} {\bibfnamefont {G.}~\bibnamefont
  {Wang}}, \bibinfo {author} {\bibfnamefont {Y.}~\bibnamefont {Song}}, \bibinfo
  {author} {\bibfnamefont {Y.}~\bibnamefont {Huang}}, \bibinfo {author}
  {\bibfnamefont {K.}~\bibnamefont {Shimada}}, \bibinfo {author} {\bibfnamefont
  {Z.}~\bibnamefont {Xu}}, \bibinfo {author} {\bibfnamefont {X.~J.}\
  \bibnamefont {Zhou}},\ and\ \bibinfo {author} {\bibfnamefont
  {G.}~\bibnamefont {Liu}},\ }\href
  {https://doi.org/10.1103/PhysRevB.103.125131} {\bibfield  {journal} {\bibinfo
   {journal} {Phys. Rev. B}\ }\textbf {\bibinfo {volume} {103}},\ \bibinfo
  {pages} {125131} (\bibinfo {year} {2021})}\BibitemShut {NoStop}%
\bibitem [{\citenamefont {Schoop}\ \emph {et~al.}(2018)\citenamefont {Schoop},
  \citenamefont {Topp}, \citenamefont {Lippmann}, \citenamefont {Orlandi},
  \citenamefont {M¨¹chler}, \citenamefont {Vergniory}, \citenamefont {Sun},
  \citenamefont {Rost}, \citenamefont {Duppel}, \citenamefont {Krivenkov},
  \citenamefont {Sheoran}, \citenamefont {Manuel}, \citenamefont {Varykhalov},
  \citenamefont {Yan}, \citenamefont {Kremer}, \citenamefont {Ast},\ and\
  \citenamefont {Lotsch}}]{2018}%
  \BibitemOpen
  \bibfield  {author} {\bibinfo {author} {\bibfnamefont {L.~M.}\ \bibnamefont
  {Schoop}}, \bibinfo {author} {\bibfnamefont {A.}~\bibnamefont {Topp}},
  \bibinfo {author} {\bibfnamefont {J.}~\bibnamefont {Lippmann}}, \bibinfo
  {author} {\bibfnamefont {F.}~\bibnamefont {Orlandi}}, \bibinfo {author}
  {\bibfnamefont {L.}~\bibnamefont {M¨¹chler}}, \bibinfo {author}
  {\bibfnamefont {M.~G.}\ \bibnamefont {Vergniory}}, \bibinfo {author}
  {\bibfnamefont {Y.}~\bibnamefont {Sun}}, \bibinfo {author} {\bibfnamefont
  {A.~W.}\ \bibnamefont {Rost}}, \bibinfo {author} {\bibfnamefont
  {V.}~\bibnamefont {Duppel}}, \bibinfo {author} {\bibfnamefont
  {M.}~\bibnamefont {Krivenkov}}, \bibinfo {author} {\bibfnamefont
  {S.}~\bibnamefont {Sheoran}}, \bibinfo {author} {\bibfnamefont
  {P.}~\bibnamefont {Manuel}}, \bibinfo {author} {\bibfnamefont
  {A.}~\bibnamefont {Varykhalov}}, \bibinfo {author} {\bibfnamefont
  {B.}~\bibnamefont {Yan}}, \bibinfo {author} {\bibfnamefont {R.~K.}\
  \bibnamefont {Kremer}}, \bibinfo {author} {\bibfnamefont {C.~R.}\
  \bibnamefont {Ast}},\ and\ \bibinfo {author} {\bibfnamefont {B.~V.}\
  \bibnamefont {Lotsch}},\ }\bibfield  {journal} {\bibinfo  {journal} {Science
  Advances}\ }\textbf {\bibinfo {volume} {4}},\ \href
  {https://doi.org/10.1126/sciadv.aar2317} {10.1126/sciadv.aar2317} (\bibinfo
  {year} {2018})\BibitemShut {NoStop}%
\bibitem [{\citenamefont {Li}\ \emph {et~al.}(2021)\citenamefont {Li},
  \citenamefont {Lv}, \citenamefont {Fang}, \citenamefont {Guo}, \citenamefont
  {Wu}, \citenamefont {Wu}, \citenamefont {Shen}, \citenamefont {Nie},
  \citenamefont {Petaccia}, \citenamefont {Cao}, \citenamefont {Xu},\ and\
  \citenamefont {Liu}}]{RN487}%
  \BibitemOpen
  \bibfield  {author} {\bibinfo {author} {\bibfnamefont {P.}~\bibnamefont
  {Li}}, \bibinfo {author} {\bibfnamefont {B.}~\bibnamefont {Lv}}, \bibinfo
  {author} {\bibfnamefont {Y.}~\bibnamefont {Fang}}, \bibinfo {author}
  {\bibfnamefont {W.}~\bibnamefont {Guo}}, \bibinfo {author} {\bibfnamefont
  {Z.}~\bibnamefont {Wu}}, \bibinfo {author} {\bibfnamefont {Y.}~\bibnamefont
  {Wu}}, \bibinfo {author} {\bibfnamefont {D.}~\bibnamefont {Shen}}, \bibinfo
  {author} {\bibfnamefont {Y.}~\bibnamefont {Nie}}, \bibinfo {author}
  {\bibfnamefont {L.}~\bibnamefont {Petaccia}}, \bibinfo {author}
  {\bibfnamefont {C.}~\bibnamefont {Cao}}, \bibinfo {author} {\bibfnamefont
  {Z.-A.}\ \bibnamefont {Xu}},\ and\ \bibinfo {author} {\bibfnamefont
  {Y.}~\bibnamefont {Liu}},\ }\href {https://doi.org/10.1007/s11433-020-1642-2}
  {\bibfield  {journal} {\bibinfo  {journal} {Science China Physics, Mechanics
  \& Astronomy}\ }\textbf {\bibinfo {volume} {64}},\ \bibinfo {pages} {237412}
  (\bibinfo {year} {2021})}\BibitemShut {NoStop}%
\bibitem [{\citenamefont {Cao}\ \emph {et~al.}(2022)\citenamefont {Cao},
  \citenamefont {Yang}, \citenamefont {Wang}, \citenamefont {Li}, \citenamefont
  {Gao}, \citenamefont {Wang}, \citenamefont {Liu}, \citenamefont {Fang},
  \citenamefont {Shi},\ and\ \citenamefont {Chen}}]{PhysRevB.106.245145}%
  \BibitemOpen
  \bibfield  {author} {\bibinfo {author} {\bibfnamefont {L.~Y.}\ \bibnamefont
  {Cao}}, \bibinfo {author} {\bibfnamefont {M.}~\bibnamefont {Yang}}, \bibinfo
  {author} {\bibfnamefont {L.}~\bibnamefont {Wang}}, \bibinfo {author}
  {\bibfnamefont {Y.}~\bibnamefont {Li}}, \bibinfo {author} {\bibfnamefont
  {B.~X.}\ \bibnamefont {Gao}}, \bibinfo {author} {\bibfnamefont
  {L.}~\bibnamefont {Wang}}, \bibinfo {author} {\bibfnamefont {J.~L.}\
  \bibnamefont {Liu}}, \bibinfo {author} {\bibfnamefont {A.~F.}\ \bibnamefont
  {Fang}}, \bibinfo {author} {\bibfnamefont {Y.~G.}\ \bibnamefont {Shi}},\ and\
  \bibinfo {author} {\bibfnamefont {R.~Y.}\ \bibnamefont {Chen}},\ }\href
  {https://doi.org/10.1103/PhysRevB.106.245145} {\bibfield  {journal} {\bibinfo
   {journal} {Phys. Rev. B}\ }\textbf {\bibinfo {volume} {106}},\ \bibinfo
  {pages} {245145} (\bibinfo {year} {2022})}\BibitemShut {NoStop}%
\bibitem [{\citenamefont {Liu}\ \emph {et~al.}(2022)\citenamefont {Liu},
  \citenamefont {Liu}, \citenamefont {Yang}, \citenamefont {Cao}, \citenamefont
  {Gao}, \citenamefont {Wang}, \citenamefont {Fang}, \citenamefont {Shi},
  \citenamefont {Yin},\ and\ \citenamefont {Chen}}]{PhysRevB.105.075111}%
  \BibitemOpen
  \bibfield  {author} {\bibinfo {author} {\bibfnamefont {J.~L.}\ \bibnamefont
  {Liu}}, \bibinfo {author} {\bibfnamefont {R.}~\bibnamefont {Liu}}, \bibinfo
  {author} {\bibfnamefont {M.}~\bibnamefont {Yang}}, \bibinfo {author}
  {\bibfnamefont {L.~Y.}\ \bibnamefont {Cao}}, \bibinfo {author} {\bibfnamefont
  {B.~X.}\ \bibnamefont {Gao}}, \bibinfo {author} {\bibfnamefont
  {L.}~\bibnamefont {Wang}}, \bibinfo {author} {\bibfnamefont {A.~F.}\
  \bibnamefont {Fang}}, \bibinfo {author} {\bibfnamefont {Y.~G.}\ \bibnamefont
  {Shi}}, \bibinfo {author} {\bibfnamefont {Z.~P.}\ \bibnamefont {Yin}},\ and\
  \bibinfo {author} {\bibfnamefont {R.~Y.}\ \bibnamefont {Chen}},\ }\href
  {https://doi.org/10.1103/PhysRevB.105.075111} {\bibfield  {journal} {\bibinfo
   {journal} {Phys. Rev. B}\ }\textbf {\bibinfo {volume} {105}},\ \bibinfo
  {pages} {075111} (\bibinfo {year} {2022})}\BibitemShut {NoStop}%
\bibitem [{\citenamefont {Demsar}\ \emph
  {et~al.}(1999{\natexlab{a}})\citenamefont {Demsar}, \citenamefont {Podobnik},
  \citenamefont {Kabanov}, \citenamefont {Wolf},\ and\ \citenamefont
  {Mihailovic}}]{PhysRevLett.82.4918}%
  \BibitemOpen
  \bibfield  {author} {\bibinfo {author} {\bibfnamefont {J.}~\bibnamefont
  {Demsar}}, \bibinfo {author} {\bibfnamefont {B.}~\bibnamefont {Podobnik}},
  \bibinfo {author} {\bibfnamefont {V.~V.}\ \bibnamefont {Kabanov}}, \bibinfo
  {author} {\bibfnamefont {T.}~\bibnamefont {Wolf}},\ and\ \bibinfo {author}
  {\bibfnamefont {D.}~\bibnamefont {Mihailovic}},\ }\href
  {https://doi.org/10.1103/PhysRevLett.82.4918} {\bibfield  {journal} {\bibinfo
   {journal} {Phys. Rev. Lett.}\ }\textbf {\bibinfo {volume} {82}},\ \bibinfo
  {pages} {4918} (\bibinfo {year} {1999}{\natexlab{a}})}\BibitemShut {NoStop}%
\bibitem [{\citenamefont {Chia}\ \emph {et~al.}(2010)\citenamefont {Chia},
  \citenamefont {Talbayev}, \citenamefont {Zhu}, \citenamefont {Yuan},
  \citenamefont {Park}, \citenamefont {Thompson}, \citenamefont {Panagopoulos},
  \citenamefont {Chen}, \citenamefont {Luo}, \citenamefont {Wang},\ and\
  \citenamefont {Taylor}}]{PhysRevLett.104.027003}%
  \BibitemOpen
  \bibfield  {author} {\bibinfo {author} {\bibfnamefont {E.~E.~M.}\
  \bibnamefont {Chia}}, \bibinfo {author} {\bibfnamefont {D.}~\bibnamefont
  {Talbayev}}, \bibinfo {author} {\bibfnamefont {J.-X.}\ \bibnamefont {Zhu}},
  \bibinfo {author} {\bibfnamefont {H.~Q.}\ \bibnamefont {Yuan}}, \bibinfo
  {author} {\bibfnamefont {T.}~\bibnamefont {Park}}, \bibinfo {author}
  {\bibfnamefont {J.~D.}\ \bibnamefont {Thompson}}, \bibinfo {author}
  {\bibfnamefont {C.}~\bibnamefont {Panagopoulos}}, \bibinfo {author}
  {\bibfnamefont {G.~F.}\ \bibnamefont {Chen}}, \bibinfo {author}
  {\bibfnamefont {J.~L.}\ \bibnamefont {Luo}}, \bibinfo {author} {\bibfnamefont
  {N.~L.}\ \bibnamefont {Wang}},\ and\ \bibinfo {author} {\bibfnamefont
  {A.~J.}\ \bibnamefont {Taylor}},\ }\href
  {https://doi.org/10.1103/PhysRevLett.104.027003} {\bibfield  {journal}
  {\bibinfo  {journal} {Phys. Rev. Lett.}\ }\textbf {\bibinfo {volume} {104}},\
  \bibinfo {pages} {027003} (\bibinfo {year} {2010})}\BibitemShut {NoStop}%
\bibitem [{\citenamefont {Chen}\ \emph {et~al.}(2014)\citenamefont {Chen},
  \citenamefont {Hu}, \citenamefont {Dong},\ and\ \citenamefont
  {Wang}}]{PhysRevB.89.075114}%
  \BibitemOpen
  \bibfield  {author} {\bibinfo {author} {\bibfnamefont {R.~Y.}\ \bibnamefont
  {Chen}}, \bibinfo {author} {\bibfnamefont {B.~F.}\ \bibnamefont {Hu}},
  \bibinfo {author} {\bibfnamefont {T.}~\bibnamefont {Dong}},\ and\ \bibinfo
  {author} {\bibfnamefont {N.~L.}\ \bibnamefont {Wang}},\ }\href
  {https://doi.org/10.1103/PhysRevB.89.075114} {\bibfield  {journal} {\bibinfo
  {journal} {Phys. Rev. B}\ }\textbf {\bibinfo {volume} {89}},\ \bibinfo
  {pages} {075114} (\bibinfo {year} {2014})}\BibitemShut {NoStop}%
\bibitem [{\citenamefont {Yusupov}\ \emph {et~al.}(2008)\citenamefont
  {Yusupov}, \citenamefont {Mertelj}, \citenamefont {Chu}, \citenamefont
  {Fisher},\ and\ \citenamefont {Mihailovic}}]{PhysRevLett.101.246402}%
  \BibitemOpen
  \bibfield  {author} {\bibinfo {author} {\bibfnamefont {R.~V.}\ \bibnamefont
  {Yusupov}}, \bibinfo {author} {\bibfnamefont {T.}~\bibnamefont {Mertelj}},
  \bibinfo {author} {\bibfnamefont {J.-H.}\ \bibnamefont {Chu}}, \bibinfo
  {author} {\bibfnamefont {I.~R.}\ \bibnamefont {Fisher}},\ and\ \bibinfo
  {author} {\bibfnamefont {D.}~\bibnamefont {Mihailovic}},\ }\href
  {https://doi.org/10.1103/PhysRevLett.101.246402} {\bibfield  {journal}
  {\bibinfo  {journal} {Phys. Rev. Lett.}\ }\textbf {\bibinfo {volume} {101}},\
  \bibinfo {pages} {246402} (\bibinfo {year} {2008})}\BibitemShut {NoStop}%
\bibitem [{\citenamefont {Albrecht}\ \emph {et~al.}(1992)\citenamefont
  {Albrecht}, \citenamefont {Kruse},\ and\ \citenamefont
  {Kurz}}]{PhysRevLett.69.1451}%
  \BibitemOpen
  \bibfield  {author} {\bibinfo {author} {\bibfnamefont {W.}~\bibnamefont
  {Albrecht}}, \bibinfo {author} {\bibfnamefont {T.}~\bibnamefont {Kruse}},\
  and\ \bibinfo {author} {\bibfnamefont {H.}~\bibnamefont {Kurz}},\ }\href
  {https://doi.org/10.1103/PhysRevLett.69.1451} {\bibfield  {journal} {\bibinfo
   {journal} {Phys. Rev. Lett.}\ }\textbf {\bibinfo {volume} {69}},\ \bibinfo
  {pages} {1451} (\bibinfo {year} {1992})}\BibitemShut {NoStop}%
\bibitem [{\citenamefont {Chen}\ \emph {et~al.}(2017)\citenamefont {Chen},
  \citenamefont {Zhang}, \citenamefont {Zhang}, \citenamefont {Dong},\ and\
  \citenamefont {Wang}}]{PhysRevLett.118.107402}%
  \BibitemOpen
  \bibfield  {author} {\bibinfo {author} {\bibfnamefont {R.~Y.}\ \bibnamefont
  {Chen}}, \bibinfo {author} {\bibfnamefont {S.~J.}\ \bibnamefont {Zhang}},
  \bibinfo {author} {\bibfnamefont {M.~Y.}\ \bibnamefont {Zhang}}, \bibinfo
  {author} {\bibfnamefont {T.}~\bibnamefont {Dong}},\ and\ \bibinfo {author}
  {\bibfnamefont {N.~L.}\ \bibnamefont {Wang}},\ }\href
  {https://doi.org/10.1103/PhysRevLett.118.107402} {\bibfield  {journal}
  {\bibinfo  {journal} {Phys. Rev. Lett.}\ }\textbf {\bibinfo {volume} {118}},\
  \bibinfo {pages} {107402} (\bibinfo {year} {2017})}\BibitemShut {NoStop}%
\bibitem [{\citenamefont {Yang}\ \emph {et~al.}(2020)\citenamefont {Yang},
  \citenamefont {Qian}, \citenamefont {Yan}, \citenamefont {Li}, \citenamefont
  {Song}, \citenamefont {Wang}, \citenamefont {Yi}, \citenamefont {Feng},
  \citenamefont {Weng},\ and\ \citenamefont {Shi}}]{PhysRevMaterials.4.094203}%
  \BibitemOpen
  \bibfield  {author} {\bibinfo {author} {\bibfnamefont {M.}~\bibnamefont
  {Yang}}, \bibinfo {author} {\bibfnamefont {Y.}~\bibnamefont {Qian}}, \bibinfo
  {author} {\bibfnamefont {D.}~\bibnamefont {Yan}}, \bibinfo {author}
  {\bibfnamefont {Y.}~\bibnamefont {Li}}, \bibinfo {author} {\bibfnamefont
  {Y.}~\bibnamefont {Song}}, \bibinfo {author} {\bibfnamefont {Z.}~\bibnamefont
  {Wang}}, \bibinfo {author} {\bibfnamefont {C.}~\bibnamefont {Yi}}, \bibinfo
  {author} {\bibfnamefont {H.~L.}\ \bibnamefont {Feng}}, \bibinfo {author}
  {\bibfnamefont {H.}~\bibnamefont {Weng}},\ and\ \bibinfo {author}
  {\bibfnamefont {Y.}~\bibnamefont {Shi}},\ }\href
  {https://doi.org/10.1103/PhysRevMaterials.4.094203} {\bibfield  {journal}
  {\bibinfo  {journal} {Phys. Rev. Materials}\ }\textbf {\bibinfo {volume}
  {4}},\ \bibinfo {pages} {094203} (\bibinfo {year} {2020})}\BibitemShut
  {NoStop}%
\bibitem [{\citenamefont {Yue}\ \emph {et~al.}(2020)\citenamefont {Yue},
  \citenamefont {Qian}, \citenamefont {Yang}, \citenamefont {Geng},
  \citenamefont {Yi}, \citenamefont {Kumar}, \citenamefont {Shimada},
  \citenamefont {Cheng}, \citenamefont {Chen}, \citenamefont {Wang},
  \citenamefont {Weng}, \citenamefont {Shi}, \citenamefont {Wu},\ and\
  \citenamefont {Feng}}]{PhysRevB.102.155109}%
  \BibitemOpen
  \bibfield  {author} {\bibinfo {author} {\bibfnamefont {S.}~\bibnamefont
  {Yue}}, \bibinfo {author} {\bibfnamefont {Y.}~\bibnamefont {Qian}}, \bibinfo
  {author} {\bibfnamefont {M.}~\bibnamefont {Yang}}, \bibinfo {author}
  {\bibfnamefont {D.}~\bibnamefont {Geng}}, \bibinfo {author} {\bibfnamefont
  {C.}~\bibnamefont {Yi}}, \bibinfo {author} {\bibfnamefont {S.}~\bibnamefont
  {Kumar}}, \bibinfo {author} {\bibfnamefont {K.}~\bibnamefont {Shimada}},
  \bibinfo {author} {\bibfnamefont {P.}~\bibnamefont {Cheng}}, \bibinfo
  {author} {\bibfnamefont {L.}~\bibnamefont {Chen}}, \bibinfo {author}
  {\bibfnamefont {Z.}~\bibnamefont {Wang}}, \bibinfo {author} {\bibfnamefont
  {H.}~\bibnamefont {Weng}}, \bibinfo {author} {\bibfnamefont {Y.}~\bibnamefont
  {Shi}}, \bibinfo {author} {\bibfnamefont {K.}~\bibnamefont {Wu}},\ and\
  \bibinfo {author} {\bibfnamefont {B.}~\bibnamefont {Feng}},\ }\href
  {https://doi.org/10.1103/PhysRevB.102.155109} {\bibfield  {journal} {\bibinfo
   {journal} {Phys. Rev. B}\ }\textbf {\bibinfo {volume} {102}},\ \bibinfo
  {pages} {155109} (\bibinfo {year} {2020})}\BibitemShut {NoStop}%
\bibitem [{\citenamefont {Rothwarf}\ and\ \citenamefont
  {Taylor}(1967)}]{PhysRevLett.19.27}%
  \BibitemOpen
  \bibfield  {author} {\bibinfo {author} {\bibfnamefont {A.}~\bibnamefont
  {Rothwarf}}\ and\ \bibinfo {author} {\bibfnamefont {B.~N.}\ \bibnamefont
  {Taylor}},\ }\href {https://doi.org/10.1103/PhysRevLett.19.27} {\bibfield
  {journal} {\bibinfo  {journal} {Phys. Rev. Lett.}\ }\textbf {\bibinfo
  {volume} {19}},\ \bibinfo {pages} {27} (\bibinfo {year} {1967})}\BibitemShut
  {NoStop}%
\bibitem [{\citenamefont {Kabanov}\ \emph {et~al.}(2005)\citenamefont
  {Kabanov}, \citenamefont {Demsar},\ and\ \citenamefont
  {Mihailovic}}]{PhysRevLett.95.147002}%
  \BibitemOpen
  \bibfield  {author} {\bibinfo {author} {\bibfnamefont {V.~V.}\ \bibnamefont
  {Kabanov}}, \bibinfo {author} {\bibfnamefont {J.}~\bibnamefont {Demsar}},\
  and\ \bibinfo {author} {\bibfnamefont {D.}~\bibnamefont {Mihailovic}},\
  }\href {https://doi.org/10.1103/PhysRevLett.95.147002} {\bibfield  {journal}
  {\bibinfo  {journal} {Phys. Rev. Lett.}\ }\textbf {\bibinfo {volume} {95}},\
  \bibinfo {pages} {147002} (\bibinfo {year} {2005})}\BibitemShut {NoStop}%
\bibitem [{\citenamefont {Chia}\ \emph {et~al.}()\citenamefont {Chia},
  \citenamefont {Zhu}, \citenamefont {Talbayev},\ and\ \citenamefont
  {Taylor}}]{RTSC}%
  \BibitemOpen
  \bibfield  {author} {\bibinfo {author} {\bibfnamefont {E.~E.~M.}\
  \bibnamefont {Chia}}, \bibinfo {author} {\bibfnamefont {J.-X.}\ \bibnamefont
  {Zhu}}, \bibinfo {author} {\bibfnamefont {D.}~\bibnamefont {Talbayev}},\ and\
  \bibinfo {author} {\bibfnamefont {A.~J.}\ \bibnamefont {Taylor}},\ }\href
  {https://doi.org/https://doi.org/10.1002/pssr.201004371} {\bibfield
  {journal} {\bibinfo  {journal} {physica status solidi (RRL) – Rapid
  Research Letters}\ }\textbf {\bibinfo {volume} {5}},\ \bibinfo {pages}
  {1}}\BibitemShut {NoStop}%
\bibitem [{\citenamefont {Lin}\ \emph {et~al.}(2020)\citenamefont {Lin},
  \citenamefont {Shi}, \citenamefont {Wang}, \citenamefont {Zhang},
  \citenamefont {Liu}, \citenamefont {Hu}, \citenamefont {Dong}, \citenamefont
  {Wu},\ and\ \citenamefont {Wang}}]{PhysRevB.101.205112}%
  \BibitemOpen
  \bibfield  {author} {\bibinfo {author} {\bibfnamefont {T.}~\bibnamefont
  {Lin}}, \bibinfo {author} {\bibfnamefont {L.~Y.}\ \bibnamefont {Shi}},
  \bibinfo {author} {\bibfnamefont {Z.~X.}\ \bibnamefont {Wang}}, \bibinfo
  {author} {\bibfnamefont {S.~J.}\ \bibnamefont {Zhang}}, \bibinfo {author}
  {\bibfnamefont {Q.~M.}\ \bibnamefont {Liu}}, \bibinfo {author} {\bibfnamefont
  {T.~C.}\ \bibnamefont {Hu}}, \bibinfo {author} {\bibfnamefont
  {T.}~\bibnamefont {Dong}}, \bibinfo {author} {\bibfnamefont {D.}~\bibnamefont
  {Wu}},\ and\ \bibinfo {author} {\bibfnamefont {N.~L.}\ \bibnamefont {Wang}},\
  }\href {https://doi.org/10.1103/PhysRevB.101.205112} {\bibfield  {journal}
  {\bibinfo  {journal} {Phys. Rev. B}\ }\textbf {\bibinfo {volume} {101}},\
  \bibinfo {pages} {205112} (\bibinfo {year} {2020})}\BibitemShut {NoStop}%
\bibitem [{\citenamefont {Chia}\ \emph
  {et~al.}(2011{\natexlab{a}})\citenamefont {Chia}, \citenamefont {Zhu},
  \citenamefont {Talbayev},\ and\ \citenamefont {Taylor}}]{Chia-2011}%
  \BibitemOpen
  \bibfield  {author} {\bibinfo {author} {\bibfnamefont {E.~E.~M.}\
  \bibnamefont {Chia}}, \bibinfo {author} {\bibfnamefont {J.-X.}\ \bibnamefont
  {Zhu}}, \bibinfo {author} {\bibfnamefont {D.}~\bibnamefont {Talbayev}},\ and\
  \bibinfo {author} {\bibfnamefont {A.~J.}\ \bibnamefont {Taylor}},\ }\href
  {https://doi.org/https://doi.org/10.1002/pssr.201004371} {\bibfield
  {journal} {\bibinfo  {journal} {physica status solidi (RRL) ¨C Rapid
  Research Letters}\ }\textbf {\bibinfo {volume} {5}},\ \bibinfo {pages} {1}
  (\bibinfo {year} {2011}{\natexlab{a}})}\BibitemShut {NoStop}%
\bibitem [{\citenamefont {Chen}\ \emph {et~al.}(2013)\citenamefont {Chen},
  \citenamefont {Dong}, \citenamefont {Wang},\ and\ \citenamefont
  {Wang}}]{LaOFeAs}%
  \BibitemOpen
  \bibfield  {author} {\bibinfo {author} {\bibfnamefont {R.}~\bibnamefont
  {Chen}}, \bibinfo {author} {\bibfnamefont {T.}~\bibnamefont {Dong}}, \bibinfo
  {author} {\bibfnamefont {H.}~\bibnamefont {Wang}},\ and\ \bibinfo {author}
  {\bibfnamefont {N.}~\bibnamefont {Wang}},\ }\href
  {https://doi.org/10.1007/s11433-013-5346-4} {\bibfield  {journal} {\bibinfo
  {journal} {Science China Physics, Mechanics and Astronomy}\ }\textbf
  {\bibinfo {volume} {56}},\ \bibinfo {pages} {2395} (\bibinfo {year}
  {2013})}\BibitemShut {NoStop}%
\bibitem [{\citenamefont {Kabanov}\ \emph {et~al.}(1999)\citenamefont
  {Kabanov}, \citenamefont {Demsar}, \citenamefont {Podobnik},\ and\
  \citenamefont {Mihailovic}}]{RT-A}%
  \BibitemOpen
  \bibfield  {author} {\bibinfo {author} {\bibfnamefont {V.~V.}\ \bibnamefont
  {Kabanov}}, \bibinfo {author} {\bibfnamefont {J.}~\bibnamefont {Demsar}},
  \bibinfo {author} {\bibfnamefont {B.}~\bibnamefont {Podobnik}},\ and\
  \bibinfo {author} {\bibfnamefont {D.}~\bibnamefont {Mihailovic}},\ }\href
  {https://doi.org/10.1103/PhysRevB.59.1497} {\bibfield  {journal} {\bibinfo
  {journal} {Phys. Rev. B}\ }\textbf {\bibinfo {volume} {59}},\ \bibinfo
  {pages} {1497} (\bibinfo {year} {1999})}\BibitemShut {NoStop}%
\bibitem [{\citenamefont {Tian}\ \emph {et~al.}(2016)\citenamefont {Tian},
  \citenamefont {Zhang}, \citenamefont {Li}, \citenamefont {Wu}, \citenamefont
  {Wu}, \citenamefont {Sun}, \citenamefont {Zhou}, \citenamefont {Wang},
  \citenamefont {Ma}, \citenamefont {Xue},\ and\ \citenamefont
  {Zhao}}]{PhysRevLett.116.107001}%
  \BibitemOpen
  \bibfield  {author} {\bibinfo {author} {\bibfnamefont {Y.~C.}\ \bibnamefont
  {Tian}}, \bibinfo {author} {\bibfnamefont {W.~H.}\ \bibnamefont {Zhang}},
  \bibinfo {author} {\bibfnamefont {F.~S.}\ \bibnamefont {Li}}, \bibinfo
  {author} {\bibfnamefont {Y.~L.}\ \bibnamefont {Wu}}, \bibinfo {author}
  {\bibfnamefont {Q.}~\bibnamefont {Wu}}, \bibinfo {author} {\bibfnamefont
  {F.}~\bibnamefont {Sun}}, \bibinfo {author} {\bibfnamefont {G.~Y.}\
  \bibnamefont {Zhou}}, \bibinfo {author} {\bibfnamefont {L.}~\bibnamefont
  {Wang}}, \bibinfo {author} {\bibfnamefont {X.}~\bibnamefont {Ma}}, \bibinfo
  {author} {\bibfnamefont {Q.-K.}\ \bibnamefont {Xue}},\ and\ \bibinfo {author}
  {\bibfnamefont {J.}~\bibnamefont {Zhao}},\ }\href
  {https://doi.org/10.1103/PhysRevLett.116.107001} {\bibfield  {journal}
  {\bibinfo  {journal} {Phys. Rev. Lett.}\ }\textbf {\bibinfo {volume} {116}},\
  \bibinfo {pages} {107001} (\bibinfo {year} {2016})}\BibitemShut {NoStop}%
\bibitem [{\citenamefont {Chia}\ \emph
  {et~al.}(2011{\natexlab{b}})\citenamefont {Chia}, \citenamefont {Zhu},
  \citenamefont {Talbayev}, \citenamefont {Lee}, \citenamefont {Hur},
  \citenamefont {Moreno}, \citenamefont {Averitt}, \citenamefont {Sarrao},\
  and\ \citenamefont {Taylor}}]{PhysRevB.84.174412}%
  \BibitemOpen
  \bibfield  {author} {\bibinfo {author} {\bibfnamefont {E.~E.~M.}\
  \bibnamefont {Chia}}, \bibinfo {author} {\bibfnamefont {J.-X.}\ \bibnamefont
  {Zhu}}, \bibinfo {author} {\bibfnamefont {D.}~\bibnamefont {Talbayev}},
  \bibinfo {author} {\bibfnamefont {H.~J.}\ \bibnamefont {Lee}}, \bibinfo
  {author} {\bibfnamefont {N.}~\bibnamefont {Hur}}, \bibinfo {author}
  {\bibfnamefont {N.~O.}\ \bibnamefont {Moreno}}, \bibinfo {author}
  {\bibfnamefont {R.~D.}\ \bibnamefont {Averitt}}, \bibinfo {author}
  {\bibfnamefont {J.~L.}\ \bibnamefont {Sarrao}},\ and\ \bibinfo {author}
  {\bibfnamefont {A.~J.}\ \bibnamefont {Taylor}},\ }\href
  {https://doi.org/10.1103/PhysRevB.84.174412} {\bibfield  {journal} {\bibinfo
  {journal} {Phys. Rev. B}\ }\textbf {\bibinfo {volume} {84}},\ \bibinfo
  {pages} {174412} (\bibinfo {year} {2011}{\natexlab{b}})}\BibitemShut
  {NoStop}%
\bibitem [{\citenamefont {Demsar}\ \emph
  {et~al.}(1999{\natexlab{b}})\citenamefont {Demsar}, \citenamefont
  {Biljakovi\ifmmode~\acute{c}\else \'{c}\fi{}},\ and\ \citenamefont
  {Mihailovic}}]{PhysRevLett.83.800}%
  \BibitemOpen
  \bibfield  {author} {\bibinfo {author} {\bibfnamefont {J.}~\bibnamefont
  {Demsar}}, \bibinfo {author} {\bibfnamefont {K.}~\bibnamefont
  {Biljakovi\ifmmode~\acute{c}\else \'{c}\fi{}}},\ and\ \bibinfo {author}
  {\bibfnamefont {D.}~\bibnamefont {Mihailovic}},\ }\href
  {https://doi.org/10.1103/PhysRevLett.83.800} {\bibfield  {journal} {\bibinfo
  {journal} {Phys. Rev. Lett.}\ }\textbf {\bibinfo {volume} {83}},\ \bibinfo
  {pages} {800} (\bibinfo {year} {1999}{\natexlab{b}})}\BibitemShut {NoStop}%
\bibitem [{\citenamefont {Hu}\ \emph {et~al.}(2014)\citenamefont {Hu},
  \citenamefont {Cheng}, \citenamefont {Yuan}, \citenamefont {Dong},\ and\
  \citenamefont {Wang}}]{PhysRevB.90.085105}%
  \BibitemOpen
  \bibfield  {author} {\bibinfo {author} {\bibfnamefont {B.~F.}\ \bibnamefont
  {Hu}}, \bibinfo {author} {\bibfnamefont {B.}~\bibnamefont {Cheng}}, \bibinfo
  {author} {\bibfnamefont {R.~H.}\ \bibnamefont {Yuan}}, \bibinfo {author}
  {\bibfnamefont {T.}~\bibnamefont {Dong}},\ and\ \bibinfo {author}
  {\bibfnamefont {N.~L.}\ \bibnamefont {Wang}},\ }\href
  {https://doi.org/10.1103/PhysRevB.90.085105} {\bibfield  {journal} {\bibinfo
  {journal} {Phys. Rev. B}\ }\textbf {\bibinfo {volume} {90}},\ \bibinfo
  {pages} {085105} (\bibinfo {year} {2014})}\BibitemShut {NoStop}%
\bibitem [{\citenamefont {Singha}\ \emph {et~al.}(2017)\citenamefont {Singha},
  \citenamefont {Pariari}, \citenamefont {Satpati},\ and\ \citenamefont
  {Mandal}}]{PhysRevB.96.245138}%
  \BibitemOpen
  \bibfield  {author} {\bibinfo {author} {\bibfnamefont {R.}~\bibnamefont
  {Singha}}, \bibinfo {author} {\bibfnamefont {A.}~\bibnamefont {Pariari}},
  \bibinfo {author} {\bibfnamefont {B.}~\bibnamefont {Satpati}},\ and\ \bibinfo
  {author} {\bibfnamefont {P.}~\bibnamefont {Mandal}},\ }\href
  {https://doi.org/10.1103/PhysRevB.96.245138} {\bibfield  {journal} {\bibinfo
  {journal} {Phys. Rev. B}\ }\textbf {\bibinfo {volume} {96}},\ \bibinfo
  {pages} {245138} (\bibinfo {year} {2017})}\BibitemShut {NoStop}%
\bibitem [{\citenamefont {Liu}\ \emph {et~al.}(2021)\citenamefont {Liu},
  \citenamefont {Zhu}, \citenamefont {Wu}, \citenamefont {Zhang}, \citenamefont
  {Song}, \citenamefont {Shi}, \citenamefont {Liu}, \citenamefont {Liu},
  \citenamefont {Song}, \citenamefont {Wu}, \citenamefont {Zhao}, \citenamefont
  {Tang}, \citenamefont {Yuan}, \citenamefont {Huang}, \citenamefont {He},
  \citenamefont {Liu}, \citenamefont {Duan},\ and\ \citenamefont
  {Meng}}]{PdTe2}%
  \BibitemOpen
  \bibfield  {author} {\bibinfo {author} {\bibfnamefont {S.-Y.}\ \bibnamefont
  {Liu}}, \bibinfo {author} {\bibfnamefont {S.-X.}\ \bibnamefont {Zhu}},
  \bibinfo {author} {\bibfnamefont {Q.-Y.}\ \bibnamefont {Wu}}, \bibinfo
  {author} {\bibfnamefont {C.}~\bibnamefont {Zhang}}, \bibinfo {author}
  {\bibfnamefont {P.-B.}\ \bibnamefont {Song}}, \bibinfo {author}
  {\bibfnamefont {Y.-G.}\ \bibnamefont {Shi}}, \bibinfo {author} {\bibfnamefont
  {H.}~\bibnamefont {Liu}}, \bibinfo {author} {\bibfnamefont {Z.-T.}\
  \bibnamefont {Liu}}, \bibinfo {author} {\bibfnamefont {J.-J.}\ \bibnamefont
  {Song}}, \bibinfo {author} {\bibfnamefont {F.-Y.}\ \bibnamefont {Wu}},
  \bibinfo {author} {\bibfnamefont {Y.-Z.}\ \bibnamefont {Zhao}}, \bibinfo
  {author} {\bibfnamefont {X.-F.}\ \bibnamefont {Tang}}, \bibinfo {author}
  {\bibfnamefont {Y.-H.}\ \bibnamefont {Yuan}}, \bibinfo {author}
  {\bibfnamefont {H.}~\bibnamefont {Huang}}, \bibinfo {author} {\bibfnamefont
  {J.}~\bibnamefont {He}}, \bibinfo {author} {\bibfnamefont {H.}~\bibnamefont
  {Liu}}, \bibinfo {author} {\bibfnamefont {Y.-X.}\ \bibnamefont {Duan}},\ and\
  \bibinfo {author} {\bibfnamefont {J.-Q.}\ \bibnamefont {Meng}},\ }\href
  {https://doi.org/https://doi.org/10.1016/j.rinp.2021.104816} {\bibfield
  {journal} {\bibinfo  {journal} {Results in Physics}\ }\textbf {\bibinfo
  {volume} {30}},\ \bibinfo {pages} {104816} (\bibinfo {year}
  {2021})}\BibitemShut {NoStop}%
\bibitem [{\citenamefont {Dai}\ \emph {et~al.}(2015)\citenamefont {Dai},
  \citenamefont {Bowlan}, \citenamefont {Li}, \citenamefont {Miao},
  \citenamefont {Wu}, \citenamefont {Kong}, \citenamefont {Richard},
  \citenamefont {Shi}, \citenamefont {Trugman}, \citenamefont {Zhu},
  \citenamefont {Ding}, \citenamefont {Taylor}, \citenamefont {Yarotski},\ and\
  \citenamefont {Prasankumar}}]{PhysRevB.92.161104}%
  \BibitemOpen
  \bibfield  {author} {\bibinfo {author} {\bibfnamefont {Y.~M.}\ \bibnamefont
  {Dai}}, \bibinfo {author} {\bibfnamefont {J.}~\bibnamefont {Bowlan}},
  \bibinfo {author} {\bibfnamefont {H.}~\bibnamefont {Li}}, \bibinfo {author}
  {\bibfnamefont {H.}~\bibnamefont {Miao}}, \bibinfo {author} {\bibfnamefont
  {S.~F.}\ \bibnamefont {Wu}}, \bibinfo {author} {\bibfnamefont {W.~D.}\
  \bibnamefont {Kong}}, \bibinfo {author} {\bibfnamefont {P.}~\bibnamefont
  {Richard}}, \bibinfo {author} {\bibfnamefont {Y.~G.}\ \bibnamefont {Shi}},
  \bibinfo {author} {\bibfnamefont {S.~A.}\ \bibnamefont {Trugman}}, \bibinfo
  {author} {\bibfnamefont {J.-X.}\ \bibnamefont {Zhu}}, \bibinfo {author}
  {\bibfnamefont {H.}~\bibnamefont {Ding}}, \bibinfo {author} {\bibfnamefont
  {A.~J.}\ \bibnamefont {Taylor}}, \bibinfo {author} {\bibfnamefont {D.~A.}\
  \bibnamefont {Yarotski}},\ and\ \bibinfo {author} {\bibfnamefont {R.~P.}\
  \bibnamefont {Prasankumar}},\ }\href
  {https://doi.org/10.1103/PhysRevB.92.161104} {\bibfield  {journal} {\bibinfo
  {journal} {Phys. Rev. B}\ }\textbf {\bibinfo {volume} {92}},\ \bibinfo
  {pages} {161104} (\bibinfo {year} {2015})}\BibitemShut {NoStop}%
\bibitem [{\citenamefont {Zhu}\ \emph {et~al.}(2021)\citenamefont {Zhu},
  \citenamefont {Zhang}, \citenamefont {Wu}, \citenamefont {Tang},
  \citenamefont {Liu}, \citenamefont {Liu}, \citenamefont {Luo}, \citenamefont
  {Song}, \citenamefont {Wu}, \citenamefont {Zhao}, \citenamefont {Liu},
  \citenamefont {Le}, \citenamefont {Lu}, \citenamefont {Ma}, \citenamefont
  {Liu}, \citenamefont {Yuan}, \citenamefont {Huang}, \citenamefont {He},
  \citenamefont {Liu}, \citenamefont {Duan},\ and\ \citenamefont
  {Meng}}]{PhysRevB.103.115108}%
  \BibitemOpen
  \bibfield  {author} {\bibinfo {author} {\bibfnamefont {S.-X.}\ \bibnamefont
  {Zhu}}, \bibinfo {author} {\bibfnamefont {C.}~\bibnamefont {Zhang}}, \bibinfo
  {author} {\bibfnamefont {Q.-Y.}\ \bibnamefont {Wu}}, \bibinfo {author}
  {\bibfnamefont {X.-F.}\ \bibnamefont {Tang}}, \bibinfo {author}
  {\bibfnamefont {H.}~\bibnamefont {Liu}}, \bibinfo {author} {\bibfnamefont
  {Z.-T.}\ \bibnamefont {Liu}}, \bibinfo {author} {\bibfnamefont
  {Y.}~\bibnamefont {Luo}}, \bibinfo {author} {\bibfnamefont {J.-J.}\
  \bibnamefont {Song}}, \bibinfo {author} {\bibfnamefont {F.-Y.}\ \bibnamefont
  {Wu}}, \bibinfo {author} {\bibfnamefont {Y.-Z.}\ \bibnamefont {Zhao}},
  \bibinfo {author} {\bibfnamefont {S.-Y.}\ \bibnamefont {Liu}}, \bibinfo
  {author} {\bibfnamefont {T.}~\bibnamefont {Le}}, \bibinfo {author}
  {\bibfnamefont {X.}~\bibnamefont {Lu}}, \bibinfo {author} {\bibfnamefont
  {H.}~\bibnamefont {Ma}}, \bibinfo {author} {\bibfnamefont {K.-H.}\
  \bibnamefont {Liu}}, \bibinfo {author} {\bibfnamefont {Y.-H.}\ \bibnamefont
  {Yuan}}, \bibinfo {author} {\bibfnamefont {H.}~\bibnamefont {Huang}},
  \bibinfo {author} {\bibfnamefont {J.}~\bibnamefont {He}}, \bibinfo {author}
  {\bibfnamefont {H.~Y.}\ \bibnamefont {Liu}}, \bibinfo {author} {\bibfnamefont
  {Y.-X.}\ \bibnamefont {Duan}},\ and\ \bibinfo {author} {\bibfnamefont
  {J.-Q.}\ \bibnamefont {Meng}},\ }\href
  {https://doi.org/10.1103/PhysRevB.103.115108} {\bibfield  {journal} {\bibinfo
   {journal} {Phys. Rev. B}\ }\textbf {\bibinfo {volume} {103}},\ \bibinfo
  {pages} {115108} (\bibinfo {year} {2021})}\BibitemShut {NoStop}%
\bibitem [{\citenamefont {Cheng}\ \emph {et~al.}(2022)\citenamefont {Cheng},
  \citenamefont {Fei}, \citenamefont {Hu}, \citenamefont {Dai}, \citenamefont
  {Song},\ and\ \citenamefont {Qi}}]{PhysRevB.106.104308}%
  \BibitemOpen
  \bibfield  {author} {\bibinfo {author} {\bibfnamefont {L.}~\bibnamefont
  {Cheng}}, \bibinfo {author} {\bibfnamefont {F.~C.}\ \bibnamefont {Fei}},
  \bibinfo {author} {\bibfnamefont {H.}~\bibnamefont {Hu}}, \bibinfo {author}
  {\bibfnamefont {Y.~M.}\ \bibnamefont {Dai}}, \bibinfo {author} {\bibfnamefont
  {F.~Q.}\ \bibnamefont {Song}},\ and\ \bibinfo {author} {\bibfnamefont
  {J.}~\bibnamefont {Qi}},\ }\href
  {https://doi.org/10.1103/PhysRevB.106.104308} {\bibfield  {journal} {\bibinfo
   {journal} {Phys. Rev. B}\ }\textbf {\bibinfo {volume} {106}},\ \bibinfo
  {pages} {104308} (\bibinfo {year} {2022})}\BibitemShut {NoStop}%
\bibitem [{\citenamefont {Lopez}(1968)}]{PhysRev.175.823}%
  \BibitemOpen
  \bibfield  {author} {\bibinfo {author} {\bibfnamefont {A.~A.}\ \bibnamefont
  {Lopez}},\ }\href {https://doi.org/10.1103/PhysRev.175.823} {\bibfield
  {journal} {\bibinfo  {journal} {Phys. Rev.}\ }\textbf {\bibinfo {volume}
  {175}},\ \bibinfo {pages} {823} (\bibinfo {year} {1968})}\BibitemShut
  {NoStop}%
\bibitem [{\citenamefont {Sheu}\ \emph {et~al.}(2013)\citenamefont {Sheu},
  \citenamefont {Chien}, \citenamefont {Uher}, \citenamefont {Fahy},\ and\
  \citenamefont {Reis}}]{PhysRevB.87.075429}%
  \BibitemOpen
  \bibfield  {author} {\bibinfo {author} {\bibfnamefont {Y.~M.}\ \bibnamefont
  {Sheu}}, \bibinfo {author} {\bibfnamefont {Y.~J.}\ \bibnamefont {Chien}},
  \bibinfo {author} {\bibfnamefont {C.}~\bibnamefont {Uher}}, \bibinfo {author}
  {\bibfnamefont {S.}~\bibnamefont {Fahy}},\ and\ \bibinfo {author}
  {\bibfnamefont {D.~A.}\ \bibnamefont {Reis}},\ }\href
  {https://doi.org/10.1103/PhysRevB.87.075429} {\bibfield  {journal} {\bibinfo
  {journal} {Phys. Rev. B}\ }\textbf {\bibinfo {volume} {87}},\ \bibinfo
  {pages} {075429} (\bibinfo {year} {2013})}\BibitemShut {NoStop}%
\bibitem [{\citenamefont {Othonos}(1998)}]{10.1063/1.367411}%
  \BibitemOpen
  \bibfield  {author} {\bibinfo {author} {\bibfnamefont {A.}~\bibnamefont
  {Othonos}},\ }\href {https://doi.org/10.1063/1.367411} {\bibfield  {journal}
  {\bibinfo  {journal} {Journal of Applied Physics}\ }\textbf {\bibinfo
  {volume} {83}},\ \bibinfo {pages} {1789} (\bibinfo {year}
  {1998})}\BibitemShut {NoStop}%
\bibitem [{\citenamefont {Vodopyanov}\ \emph {et~al.}(1992)\citenamefont
  {Vodopyanov}, \citenamefont {Graener}, \citenamefont {Phillips},\ and\
  \citenamefont {Tate}}]{PhysRevB.46.13194}%
  \BibitemOpen
  \bibfield  {author} {\bibinfo {author} {\bibfnamefont {K.~L.}\ \bibnamefont
  {Vodopyanov}}, \bibinfo {author} {\bibfnamefont {H.}~\bibnamefont {Graener}},
  \bibinfo {author} {\bibfnamefont {C.~C.}\ \bibnamefont {Phillips}},\ and\
  \bibinfo {author} {\bibfnamefont {T.~J.}\ \bibnamefont {Tate}},\ }\href
  {https://doi.org/10.1103/PhysRevB.46.13194} {\bibfield  {journal} {\bibinfo
  {journal} {Phys. Rev. B}\ }\textbf {\bibinfo {volume} {46}},\ \bibinfo
  {pages} {13194} (\bibinfo {year} {1992})}\BibitemShut {NoStop}%
\bibitem [{\citenamefont {Lv}\ \emph {et~al.}(2019)\citenamefont {Lv},
  \citenamefont {Chen}, \citenamefont {Qiao}, \citenamefont {Ma}, \citenamefont
  {Yang}, \citenamefont {Li}, \citenamefont {Wang}, \citenamefont {Tao},\ and\
  \citenamefont {Xu}}]{Lv_2019}%
  \BibitemOpen
  \bibfield  {author} {\bibinfo {author} {\bibfnamefont {B.}~\bibnamefont
  {Lv}}, \bibinfo {author} {\bibfnamefont {J.}~\bibnamefont {Chen}}, \bibinfo
  {author} {\bibfnamefont {L.}~\bibnamefont {Qiao}}, \bibinfo {author}
  {\bibfnamefont {J.}~\bibnamefont {Ma}}, \bibinfo {author} {\bibfnamefont
  {X.}~\bibnamefont {Yang}}, \bibinfo {author} {\bibfnamefont {M.}~\bibnamefont
  {Li}}, \bibinfo {author} {\bibfnamefont {M.}~\bibnamefont {Wang}}, \bibinfo
  {author} {\bibfnamefont {Q.}~\bibnamefont {Tao}},\ and\ \bibinfo {author}
  {\bibfnamefont {Z.-A.}\ \bibnamefont {Xu}},\ }\href
  {https://doi.org/10.1088/1361-648x/ab2498} {\bibfield  {journal} {\bibinfo
  {journal} {Journal of Physics: Condensed Matter}\ }\textbf {\bibinfo {volume}
  {31}},\ \bibinfo {pages} {355601} (\bibinfo {year} {2019})}\BibitemShut
  {NoStop}%
\end{thebibliography}%

\end{document}